\documentclass[longauth]{aa}  
\usepackage{graphicx}
\usepackage{caption}
\usepackage{subcaption}
\usepackage{xcolor}
\usepackage{soul}
\usepackage{dblfloatfix}
\usepackage{txfonts}
\usepackage{mathabx}
\usepackage[hidelinks]{hyperref}
\usepackage[normalem]{ulem}
\usepackage{orcidlink}
\usepackage{booktabs}
\usepackage{verbatim}
\usepackage{siunitx}
\usepackage{textcomp}
\usepackage{amsmath}
\usepackage{amsfonts}
\usepackage{amssymb}
\usepackage{float}

\newcounter{pta}

\DeclareMathSizes{12}{30}{16}{12}
\usepackage{amsmath,scalerel}

\makeatletter
\renewcommand*\aa@pageof{, page \thepage{} of \pageref*{LastPage}}
\makeatother

\usepackage{natbib}
\setcitestyle{authoryear,open={(},close={)}} 

\usepackage{titlesec}

\titlespacing\section{0pt}{14pt plus 2pt minus 2pt}{5pt plus 1pt minus 1pt}
\titlespacing\subsection{0pt}{12pt plus 4pt minus 2pt}{6pt plus 2pt minus 2pt}
\titlespacing\subsubsection{0pt}{12pt plus 4pt minus 2pt}{6pt plus 2pt minus 2pt}

\usepackage[switch]{lineno} 

\begin{document} 

 \title{Exploring the time variability of the Solar Wind using LOFAR pulsar data}

\author{S.~C.~Susarla\orcidlink{0000-0003-4332-8201}\inst{\ref{uog}}, 
        A.~Chalumeau\orcidlink{https://orcid.org/0000-0003-2111-1001}\inst{\ref{unimib},\ref{infn-unimib}},
        C.~Tiburzi\orcidlink{0000-0001-6651-4811}\inst{\ref{inaf-oac}}, 
        E.~F.~Keane\orcidlink{0000-0002-4553-655X}\inst{\ref{tcd}},
        J.~P.~W.~Verbiest\orcidlink{0000-0002-4088-896X}\inst{\ref{FSI}},
        {J.~S.~Hazboun\orcidlink{0000-0003-2742-3321}\inst{\ref{osu}}},
        M.~A.~Krishnakumar\orcidlink{0000-0003-4528-2745}\inst{\ref{mpifr},\ref{unibi},\ref{ncra}},
        F. Iraci\orcidlink{0009-0001-0068-4727}\inst{\ref{inaf-oac}, \ref{UniCA}},
		G.~M.~Shaifullah\orcidlink{0000-0002-8452-4834}\inst{\ref{unimib},\ref{infn-unimib},\ref{inaf-oac}},
        A.~Golden\orcidlink{0000-0001-8208-4292}\inst{\ref{uog}},
        A.-S.~Bak~Nielsen\orcidlink{0000-0002-1298-9392}\inst{\ref{mpifr}},
        J.~Donner\inst{\ref{mpifr},\ref{unibi}},
        J.-M.~Grie{\ss}meier\orcidlink{0000-0003-3362-7996}\inst{\ref{lpc2e},\ref{orleans}},
        M.~J.~Keith\orcidlink{0000-0001-5567-5492}\inst{\ref{jbca}},
        S.~Os{\l}owski\orcidlink{0000-0003-0289-0732}\inst{\ref{atnf}},
        N.~K.~Porayko\orcidlink{0000-0002-6955-8040}\inst{\ref{unimib}, \ref{mpifr}},
        M.~Serylak\orcidlink{0000-0002-6670-652X}\inst{\ref{skauk},\ref{uwcsa}},
        J.~M.~Anderson\orcidlink{0000-0002-5989-8498}\inst{\ref{atbp}}
        M.~Br\"uggen\orcidlink{0000-0002-3369-7735}\inst{\ref{uniha}},
        B.~Ciardi\orcidlink{0000-0002-5037-310X}\inst{\ref{mpiag}},
        R.-J.~Dettmar\orcidlink{0000-0001-8206-5956}\inst{\ref{rub}},
        M.~Hoeft\orcidlink{0000-0001-5571-1369}\inst{\ref{tauten}},
        J.~K\"unsem\"oller\inst{\ref{unibi}},
        D.~Schwarz\orcidlink{0000-0003-2413-088}\inst{\ref{unibi}},
        C.~Vocks\orcidlink{0000-0001-8583-8619}\inst{\ref{aip}}
        }
    
\institute{
Physics, School of Natural Sciences, Ollscoil na Gaillimhe --- University of Galway, University Road, Galway, H91 TK33, Ireland\label{uog}\and
Dipartimento di Fisica ``G. Occhialini", Universit{\'a} degli Studi di Milano-Bicocca, Piazza della Scienza 3, I-20126 Milano, Italy\label{unimib}\and
INFN, Sezione di Milano-Bicocca, Piazza della Scienza 3, I-20126 Milano, Italy\label{infn-unimib}\and
INAF - Osservatorio Astronomico di Cagliari, via della Scienza 5, 09047 Selargius (CA), Italy\label{inaf-oac}\and
School of Physics, Trinity College Dublin, College Green, Dublin 2, D02 PN40, Ireland\label{tcd}\and
Florida Space Institute, University of Central Florida, 12354 Research Parkway, Partnership 1 Building, Suite 214, Orlando, 32826-0650, FL, USA\label{FSI}\and
Oregon State University, 1500 SW Jefferson Ave, Corvallis, OR 97331, United States\label{osu}\and
Max-Planck-Institut f{\"u}r Radioastronomie, Auf dem H{\"u}gel 69, 53121 Bonn, Germany\label{mpifr}\and
Fakult{\"a}t f{\"u}r Physik, Universit{\"a}t Bielefeld, Postfach 100131, 33501 Bielefeld, Germany\label{unibi}\and
 National Centre for Radio Astrophysics, Pune University Campus, Pune 411007, India \label{ncra}\and
 Dipartimento di Fisica, Università di Cagliari, Cittadella Universitaria, I-09042 Monserrato (CA), Italy \label{UniCA}\and
 LPC2E - Universit\'{e} d'Orl\'{e}ans /  CNRS, 45071 Orl\'{e}ans cedex 2, France\label{lpc2e}\and
 Observatoire Radioastronomique de Nan\c{c}ay (ORN), Observatoire de Paris, Universit\'{e} PSL, Univ Orl\'{e}ans, CNRS, 18330 Nan\c{c}ay, France \label{orleans}\and
 Jodrell Bank Centre for Astrophysics, Department of Physics and Astronomy, University of Manchester, Manchester M13 9PL, UK\label{jbca}\and
  Australia Telescope National Facility, CSIRO, Space and Astronomy, PO Box 76, Epping, NSW 1710, Australia\label{atnf}\and
  SKA Observatory, Jodrell Bank, Lower Withington, Macclesfield, SK11 9FT, United Kingdom\label{skauk}\and
 Department of Physics and Astronomy, University of the Western Cape, Bellville, Cape Town, 7535, South Africa\label{uwcsa}\and
 Leibniz Institute for Agricultural Engineering and Bioeconomy (ATB), Max-Eyth-Allee 100, 14469 Potsdam, Germany\label{atbp}\and
University of Hamburg, Gojenbergsweg 112, 21029 Hamburg, Germany \label{uniha}\and
 Max Planck Institute for Astrophysics, Karl-Schwarzschild-Str. 1, 85741 Garching, Germany\label{mpiag}\and 
 Ruhr University Bochum, Faculty of Physics and Astronomy, Astronomical Institute (AIRUB), 44780 Bochum, Germany\label{rub}\and
 Thüringer Landessternwarte Tautenburg, Sternwarte 507778 Tautenburg,Germany\label{tauten}\and
  Leibniz Institute for Astrophysics Potsdam (AIP)\label{aip}
} 

   \date{Received XXX; accepted YYY}
\authorrunning{LOFAR}

 
  \abstract
{
High-precision pulsar timing is highly dependent on precise and accurate modeling of any effects that impact the data. In particular, effects that contain stochastic elements contribute to corruption and complexity in the analysis of pulsar-timing data. It was shown that commonly used Solar Wind models do not accurately account for variability in the amplitude of the Solar wind on both short and long time scales.}
{
In this study, we test and validate a new, cutting-edge Solar wind modeling method included in the \texttt{enterprise} software suite (widely used for pulsar noise analysis) through extended simulations, and we apply it to investigate temporal variability in LOFAR data.  
Our model testing scheme in itself provides an invaluable asset for pulsar timing array (PTA) experiments. As improperly accounting for the solar wind signature in pulsar data can induce false-positive signals, it is of fundamental importance to include in any such investigations.
}
{
We employ a Bayesian approach utilizing a continuously varying Gaussian process to model the solar wind. It uses a spherical approximation that modulates the electron density. This method, which we refer to as a Solar Wind Gaussian Process (SWGP), is integrated into existing noise analysis software, specifically \texttt{enterprise}. Validation of this model is performed through simulations. We then conduct noise analysis on eight pulsars from the LOFAR dataset with most pulsars having a timespan of $\sim 11$ years encompassing one full solar activity cycle. Furthermore, we derive the electron densities from the dispersion measure values obtained by the SWGP model. 
}
{
Our analysis reveals a strong correlation between the electron density at 1 AU and the ecliptic latitude (ELAT) of the pulsar. Pulsars with $|ELAT|< 3^{\circ}$ exhibit significantly higher average electron densities. Furthermore, we observe distinct temporal patterns in electron densities in different pulsars. In particular, pulsars within $|ELAT|< 3^{\circ}$ exhibit similar temporal variations, while the electron densities of those outside this range correlate with the solar activity cycle. Notably, some pulsars exhibit sensitivity to the solar wind up to $45^{\circ}$ away from the Sun in LOFAR data.
}
{
The continuous variability in electron density offered in this model represents a substantial improvement over previous models, which assume a single value for piece-wise bins of time. This advancement holds promise for solar wind modeling in future International Pulsar Timing Array data combinations. }


   \maketitle
%
\section{Introduction}
\label{introduction}
The solar wind (SW) is a highly magnetized stream of plasma propagating in interplanetary space from the hot solar corona, first mentioned in \citet{Biermann1951}. The composition of the SW is a mixture of materials found in the solar plasma, composed of ionized hydrogen (electrons and protons) with an 8\% component of ionized helium (also called $\alpha$ particles) and trace amounts of heavy ions and atomic nuclei (e.g. \citet{Feldman1998}). The Ulysses spacecraft \citep{marsdenulysses}, with its near-polar orbit, has revealed that the SW  exists in a bimodal state: an irregular and dense slow wind with typical speeds of $\sim$400 km/s and a smooth fast wind with a speed of $\sim$750 km/s \citep{ulysses}. This fast wind has been shown to originate from coronal holes and the slow wind from the boundaries or interiors of streamers (see e.g. Figure 1 in \citealt{tiburzi2019}).\\

Pulsars are rapidly rotating neutron stars mainly visible as regularly pulsating radio sources.
Their rotation is so reliable that it can be used as a highly precise clock-like signal. By studying
this clock signal, via their emitted radio pulses, pulsars can be used to probe several effects,
such as interstellar weather associated with the SW. This is done by measuring the electron content of the heliosphere, due to SW-induced modifications on a given pulsar's transiting radio pulse, also known as \textit{dispersion}. These modifications are quantified using a metric defined as the Dispersion Measure (DM), i.e. the integral of the column density of free electrons $n^{LoS}_e$ along the line-of-sight (LoS):

\begin{equation}
    \mathrm{DM} = \int_{\rm LoS} n^{LoS}_e dl.
    \label{dmeq}
\end{equation}
Dispersion causes a delay in the arrival of pulsar emission that depends on the frequency of the radiation and the DM parameter (measured in $\rm pc\ \rm cm^{-3}$). This delay can be written as:

\begin{equation}
    \Delta t = \frac{\mathrm{DM}}{K_D \nu^2}\;,
    \label{timedel}
\end{equation}
where $K_D \simeq 2.41 \times 10^{-4} \rm MHz^{-2} \rm pc\ \rm cm^{-3} s^{-1}$ is the dispersion constant and $\nu$ is the observing frequency \citep{kulkarni2020}. Equation~\ref{dmeq} shows that the DM can vary because of changes in electron density along a given LoS, and these changes can be tracked thanks to the inverse-square dependency on the observing frequency, with pulsar observations collected with low-frequency facilities being particularly sensitive to this effect. The measured DM has components related to both the SW and changes in the Ionized Interstellar Medium (IISM), which have differing time-varying signatures that can be used to disentangle the two components.\\

The DM parameter can be calculated via a process called \textit{pulsar timing} (e.g. \citealt{LorimerandKramer2005}). This involves monitoring the arrival times of radio pulses from a pulsar at an observatory. These Times of Arrival (ToAs) are converted to solar system barycentric arrival times for analysis in an inertial reference frame. A mathematical model based on an ensemble of pulsar characteristics, also known as the Timing Model (TM), is then used to compare and quantify factors affecting the ToAs. This technique enables precise measurement of the pulsar parameters on which the TM is based, with accuracy increasing with longer data sets and improved ToA measurements. Different phenomena introduce noise in the timing residuals (the difference between the observed ToAs and the arrival times predicted by the TM), such as gravitational waves (GWs). Some of these noise sources can be accurately characterized by measuring the correlated signatures in the residuals in an array of pulsars. This is done by using decades of observations of several millisecond pulsars (MSPs, \citealt{mspcitation}) observed using different telescopes. This experimental methodology forms the basis for a pulsar timing array (PTA). Several PTA collaborations like the European Pulsar Timing Array (EPTA, \citealt{eptadr2}), North American Nanohertz Observatory for Gravitational Waves (NANOGrav, \citealt{nanograv15y}), Parkes Pulsar Timing Array (PPTA, \citealt{pptadr3}) and Indian Pulsar Timing Array (InPTA, \citealt{inptadr1}) are combining their datasets to form a global consortium called International Pulsar Timing Array (IPTA) to have a clear detection of the Gravitational Wave Background (GWB, \citealt{hd1983}). In 2023, PTAs like the EPTA+InPTA, NANOGrav and the PPTA collaborations (see e.g. \citealt{eptagwb,nanogravgwb,pptagwb}) identified a correlated signature consistent with a GWB at the level of at least 3$\sigma$ confidence. A variety of noise sources can induce a false detection of the GWB. One such noise source in PTAs is the SW as shown by \citet{Tiburzi2015}, hence it is essential to fully understand its contribution to the TM solutions to optimise recovery of any underlying GWB signals.\\

Several studies have been carried out to observe the SW using pulsars (e.g. \citealt{counselman1972,madison2019, Tokumaru2020, Tiburzi2021,kumar2022}). It is worth noting that SW can change the DM by a measure of approximately 0.01 $\rm pc\ \rm cm^{-3}$ for the MSPs whose LoS gets close to the Sun. In reality, for most pulsars it is one order of magnitude smaller and these contributions become important as they are time variable and the current precision of the DMs achieved in the low-frequency observation is much smaller than the DM contribution by SW (see for eg. \citet{Donner202036msp}). The standard approach in pulsar timing is to model the SW as a time-independent spherical distribution of free electrons, where the DM varies according to the following equation:

\begin{equation}
    DM_{\rm{SW}} = n_{\rm e}\frac{\rho}{r_e \sin\rho}[1\mathrm{AU}]^2,
    \label{spherical}
\end{equation}

where $\rho$ is the pulsar-Sun-observer angle, $n_{\rm e}$ is the electron density at 1 AU and $r_e$ is the distance between the observatory and the Sun \citep{tiburzi2019}. \citet{You2007} proposed an alternative model that took into account the bimodal nature of the SW. They used distinct radial distributions of free electrons for each of the two streams within the SW, fast and slow, and utilized solar magnetograms obtained from the Wilcox observatory\footnote{http://wso.stanford.edu/} to differentiate the LoS components that were influenced by one or the other stream. The total contribution of the SW was obtained by adding these individual contributions. \\
Both of these models have shortcomings, as demonstrated by  
\citet{tiburzi2019}, who formally compared the performance of the two approaches, in addition to adding a time-variable amplitude to the spherical model, using low-frequency observations of the binary radio MSP PSR~J0034$-$0534. This work showed that neither model provided a satisfactory description of the SW effects on the dataset, although the spherical model performed systematically better than the bimodal one, in contrast to the conclusions of \citet{You2007}. The observed inconsistency between the \citet{You2007} and \citet{tiburzi2019} analyses is believed to stem from either of the following: (i) the enhanced precision in DM reached thanks to the lower observing frequency in the dataset utilized by \citet{tiburzi2019} in contrast to that employed by \citet{You2007}; or (ii) a potential difference in the effectiveness of the two-phase model concerning the heliospheric latitude of the pulsar, as both the studies examined data from different pulsars. 
In the same year, \citep{madison2019} found the optimal value of $n_{\rm e}$ to be $7.9\pm 0.2$~$\mathrm{cm}^{-3}$ via an analysis of the NANOGrav 11-yr dataset \citep{11yrnano}. However, it was noted that this value could be significantly improved if one used a lower observing frequency and larger temporal baseline. \citet{Tiburzi2021} subsequently showed that a time-variable SW amplitude is a better description of the SW signal in pulsar data, compared to a constant one as previously used. Working in the context of PTAs, \citet{Hazboun2022} demonstrated that by relaxing the assumption that the electron density around the Sun drops off as $1/r^2$, and by assuming a more general electron density decreasing as $1/r^\gamma$, improved results could be achieved. They also compared the use of a piece-wise binned time dependence for $n_e$ to a deterministic Fourier-basis model, both used in order to more easily model the time dependence across multiple pulsars. A recent study by \citet{luliana2024} has examined the effectiveness of Gaussian processes on estimating the effect the SW has on pulsar pulse dispersion. The authors showed that fitting for a piece-wise function for each solar conjunction using Gaussian Processes effectively encapsulates the variability associated with the SW. Whilst these authors presented in conclusion an annual single value fit for the $n_{e}$, we note that this does not take into account the continuous variability of SW electron densities which has been observed by numerous 'in situ' spacecraft operating in the inner solar system.\\

Ongoing efforts to incorporate low-frequency radio data from the LOw-Frequency ARray (LOFAR, van Haarlem et al 2013) telescopes into the upcoming IPTA data release are underway. Therefore, modelling the impact of the SW-associated dispersion on TMs is of key significance as their effect is strongly pronounced in this data set, making it imperative to model this noise source accurately and formally integrating a functioning SW noise model into the existing noise analysis packages, e.g., \texttt{enterprise} \citep{enterprise}.\\

In this study, we present \textit{Solar Wind Gaussian Process} (SWGP), a method developed and tested by the authors, and integrated into \texttt{enterprise\_extensions} \citep{enterprise} to account for the time-variability of the dispersion associated with the SW. SWGP uses Bayesian analysis techniques to examine the SW structure within the context of pulsar timing. Initially, we describe how simulations are used to assess the performance of the model. After this tuning phase, we then apply the model to pulsar data obtained with LOFAR. A total of eight pulsars are considered in our investigation, numerically increasing the sample used in \citet{luliana2024} in which only three pulsars were considered. We demonstrate the improvement of this approach over previous methods, such as that employed by \citet{Tiburzi2021}. Furthermore, we explore evidence for SW-associated variability with changing ecliptic latitudes, and we also compare the derived $n_{\rm e}$ values obtained from the pulsars with 'in situ' measurements from spacecraft. We confirm that the study of the heliospheric SW using radio pulsars is capable of resolving the two-phase structure of SW, confirmed by data obtained from the Ulysses mission. We also discuss the viability of considering the SW contribution as a common noise source when performing the TM analyses as part of the formal IPTA analysis workflow to detect GWB signatures.

This paper is divided into the following sections. In \S\ref{model}, we describe the model and provide its theoretical background and context. In \S\ref{simulations}, we show the performance of the described model on simulations. We describe the pulsar timing dataset, and how it was obtained, in \S\ref{LOFARdata}. Our modelling methods are applied to these data in \S\ref{applicationswgp}, where we also discuss the results. In \S\ref{commonsec} we then test the viability of using SW as a common noise before concluding in \S\ref{conclusions}.

\section{Methodology \& Tools} 
\label{model}

In this section, we briefly describe the models that we used for each noise process.  All the models described in this section have been incorporated in the \texttt{enterprise} software suite \citep{enterpriseellis}. We explain the Bayesian framework that is used in this paper and the definitions of Gaussian likelihood, in addition to provide a theoretical basis for each of the models used as part of the data analysis in this study.

\subsection{Bayesian framework and Likelihood}

The methodology of this paper is primarily based on the Bayesian approach to parameter inference. The physical imprints that are embedded in the timing residuals can be characterized by several parameters. All these parameters are considered as random variables, and their associated probability distribution functions are calculated according to the Bayes theorem for each of those parameters. In our case, we assume that the noise in the timing residuals are characterised by various parameters that are listed in Table \ref{tab:prior_table}. The priors of each of the parameters are sampled according to Gaussian Processes (GP). GPs are used to model stochastic variations such that every finite collection of the random variables follow a multivariate normal distribution \citep{rasmussen}.

The timing residuals ($\delta t$) contain two kinds of components namely, stochastic and deterministic. The Gaussian likelihood for the timing residuals can be defined in the time domain as:
\begin{equation}
\begin{split}
\label{eq:gwb_reduc_lik}
L&(\delta t|\theta_d,\theta_s) =\\ 
&\frac{\rm{exp}\left[-\frac{1}{2}\sum_{ij}\bigg(\delta t_i - \mathcal{D}(t_i;\theta_d)\bigg)^{T} \Vec{C}_{ij}^{-1}(\theta_s)\bigg(\delta t_j - \mathcal{D}(t_j;\theta_d)\bigg)\right]}{\sqrt{(2\pi)^n|\Vec{C}|}} ,
\end{split}
\end{equation}
where $i,j \in  \{1,\,2,\,\ldots,\,n\}$, $n$ being the number of ToAs. $\mathcal{D}$ is the time domain function representing any deterministic signal, $\Vec{C}$ is the time domain covariance matrix, where the stochastic signals are included. They are parameterised by $\theta_d$ and $\theta_s$ respectively \citep{vanhaasteren2009}. In this work, $\mathcal{D}$ corresponds to the deterministic parameter, i.e., electron density at 1 AU from the SW which is represented by $\widebar{n}_e$. More details on the deterministic component are presented in \S\ref{deter}.

The general covariance matrix is decomposed into the following stochastic components:
\begin{equation}
\label{eq:covariance}
\Vec{C} = \Vec{C}_{\rm{TMe}} + \Vec{C}_{\rm{WN}} + \Vec{C}_{\rm{DM,ISM}} + \Vec{C}_{\rm{RN}} + \Vec{C}_{\rm{SW}},
\end{equation}
where each term represents the covariance matrix corresponding to Timing Model errors, White Noise, DM noise, Red Noise and time variable SW respectively. All of these noise components are described in the following subsections.

\subsection{Timing model marginalization}
\label{tmm}
In conventional practice, the assumption of the best-fit timing solution only comprising radiometer noise tends to overfit the overall solution which can introduce bias towards other unmodeled sources. To address this, fitting all parameters in the timing model as Bayesian hyperparameters within the {\tt enterprise} package has been considered. However, this method is inefficient as it can be computationally expensive. An alternative approach involves analytically marginalizing the likelihood over the errors associated with timing model parameters \citep{chalumeau2021}. It has been demonstrated by \citet{van_Haasteren_2014} that this process is equivalent to the marginalization of a corresponding Gaussian process with an improper prior. The covariance matrix for timing model errors can be defined as follows:
\begin{equation}
    \Vec{C}_{\rm{TMe}}(t_i,t_j) = \sum_{k,l}^N\ M_k(t_i)\Sigma_{kl}M_l(t_j),
\end{equation}

where $M$ is the design matrix which contains the partial derivatives of the timing residuals with respect to the timing model parameters, and $\Sigma=\lambda I$ where $I$ is the identity matrix and $\lambda$ is a large numerical constant. Note that the covariance matrix $\Vec{C}_{\rm{TMe}}$ does not contain any parameter to fit for. It is a way to marginalize over the TM errors which inherently assumes a linear model of all the parameters in the TM. 

\subsection{White noise}
\label{wn}

A predominant element observed in pulsar timing data is white noise (WN), distinguished by its stochastic fluctuations and the absence of obvious periodic patterns. This is generally caused by the radiometer noise from the instruments and from pulse jitter noise (\citealt{liujitter,wang2015}). The ToAs are calculated using cross-correlation between the template profile and the integrated pulse profile at distinct epochs. Due to the presence of WN, we can redefine the uncertainties in the ToAs that are quantified by their ToA errors ($\sigma_{ToA}$), which can be adjusted as follows:
\begin{equation}
    \sigma = \sqrt{E_f^2\sigma_{ToA}^2+E_q^2}.
\end{equation}

Here, $\sigma$ represents the new uncertainty after accounting for WN, $E_f$ or EFAC is the multiplicative factor that is attributed to the uncertainty related to radiometer noise, while $E_q$ or EQUAD accounts for various stochastic noises such as profile variations by adding in quadrature. As a result, the WN covariance matrix, ($\Vec{C}_{\rm{WN}}$) is given by:
\begin{equation}
    \Vec{C}_{\rm{WN}} = (E_f^2\sigma_{ToA}^2(t_i)+E_q^2)\delta_{ij}
\end{equation}

where $i$ and $j$ indices denote corresponding backend's ToAs. $\delta_{ij}$ representing the Kronecker delta function; EFAC and EQUAD serve as empirical parameters characterizing white noise for each system \citep{chalumeau2021}.

\subsection{Red signals}
\label{rn}
Red signals are noise processes with an associated ``red'' power spectrum, i.e., dominated by low frequencies. It is essential to model such red-noise signals, as they can have similar signature as that of nano-hertz gravitational waves within PTA timing solution data. A study by \citet{hazboun2020} clearly demonstrated the need to model these signals accurately to ensure recovery of valid GWB detections. Similar to \citet{chalumeau2021}, we adopt a function-space view of the Gaussian process. The timing residuals due to the red signals at each epoch $t_i$ are modeled as:
\begin{equation}
    \delta t (t_i) = \sum_{l=1}^N X_l F_{2l-1}(t_i)+Y_l F_{2l}(t_i),
\end{equation}

where $X_l$ and $Y_l$ play the role of weights and the fourier basis functions are:
\begin{equation}
\label{frs}
\begin{aligned}
    F_{2l-1}(t_i) = \cos(2\pi t_if_l),\\
    F_{2l}(t_i) = \sin(2\pi t_if_l),
\end{aligned}
\end{equation}

where $l=1,\,2,\,\ldots,\,N$. Here, $N$ represents the number of Fourier components. This representation aligns with the conventional Fourier transform under the condition that $f=l/T$ (where $T$ denotes the total timespan) in the scenario of evenly spaced observing epochs, $t_i$. However, in our dataset, observations tend to be irregular with an uneven cadence, leading to the non-orthogonality of the Fourier bases. Nonetheless, for this study, we adopt a set of evenly spaced frequencies $\Delta f = 1/T$, commencing at $f=1/T$ and terminating at $N/T$. 

The covariance matrix $\mathbf{\Sigma}$ governing the Fourier coefficients is determined by the power spectral density (PSD) $S(f)$, with the simplest model being a power law:
\begin{equation}
\label{psdpowerlaw}
    S_P(f;\theta_s) = \frac{A^2}{12\pi^2}\left(\frac{f}{yr^{-1}}\right)^{-\gamma} yr^3,
\end{equation}

described with parameters $\theta_s = (A, \gamma)$, where the amplitude $A$ is the normalised value at the frequency of $1/1yr$, and $\gamma$ is the spectral index. The covariance matrix in the frequency domain is thus given by:
\begin{equation}
\label{covarfreq}
    \Sigma_{kl}(\theta_s) = S_P(f_k;\theta_s) \delta_{kl}/T,
\end{equation}

where $k,l = 1,2,....,N$. For our purposes, we consider only spatially uncorrelated red signals. 

\subsubsection{Achromatic red noise}
\label{arn}
Achromatic red noise (RN) is widely used in single-pulsar noise models to depict the long-term variation of the pulsar spin. Also known as `timing noise' or `spin noise', RN significantly affects ToAs in younger pulsars, and various physical mechanisms have been proposed to explain it, including magnetospheric variability and interactions between the superfluid core of the pulsar and the solid crust (\citealt{alparptn,Hobbs06ptn}). The origins of RN in MSPs may differ from those in young pulsars due to their substantially weaker magnetic fields; superfluid turbulence in the stellar interior has been suggested as a potential contributor to RN in MSPs (\citet{superflu}). We model the RN according to the description given above. This noise component is unique for each pulsar. Furthermore, it is not radio-frequency dependent, and is spatially uncorrelated among different pulsars. The RN covariance matrix is: 
\begin{equation}
    \Vec{C}_{\rm{RN}}(t_i,t_j;\theta_s) = \sum_{k,l}^N\ F^{RN}_k(t_i)\Sigma^{RN}_{kl}(\theta_s)F^{RN}_l(t_j),
\end{equation}

where $F^{RN}$ is the Fourier basis functions as they are detailed in equation \ref{frs} and $\Sigma^{RN}$ is the covariance matrix in the frequency domain. 

\subsubsection{Chromatic red noise}
\label{crn}
Chromatic red noise or DM noise is caused by the plasma along the pulsar's line of sight (LoS), including contributions from the interstellar medium (ISM), the solar system interplanetary medium, and the terrestrial ionosphere. As the pulsar signal travels to the observatory, it gets dispersed by all the plasma along the LoS, which causes a time delay scaling as $\Delta t\ \propto\ \nu^{-2}$. The most prominent contribution to this dispersive delay typically comes from the ISM. As mentioned, this time delay is quantified by the DM parameter, that can vary because of variations in the LoS, causing time-correlated noise in the timing residuals - such variability needs to be quantified so as not to obscure GWB detections (\citet{keith2013},\citet{you2007dm}). We use a similar model to describe the DM variations (henceforth known as DMv) as for the achromatic red noise. 

We model DM noise as a power law with parameters $A_{DM}$ and $\gamma_{DM}$, with Fourier basis components $F^{DM}\ \propto\ \kappa_j$ , where $\kappa_j = K^2\nu_j^{-2}$ is introduced to model the dependence of DM noise amplitude on the radio frequency $\nu_j$ of ToA $j$~\citep{van_Haasteren_2014}. We choose $K = 1400$~MHz to be the reference frequency in our case. The covariance matrix for DM noise can be detailed as:
\begin{equation}
    \Vec{C}_{\rm{DM},\rm{ISM}}(t_i,t_j;\theta_s) = \sum_{k,l}^N\ F^{DM}_k(t_i)\Sigma^{DM,ISM}_{kl}(\theta_s)F^{DM}_l(t_j),
\end{equation}

where the $F^{DM}$ represent the Fourier basis functions of the DM noise and $\Sigma^{DM,ISM}$ is the covariance matrix in the frequency domain for DM noise, which is detailed in equation \ref{covarfreq}. Please note that the presence of other chromatic effects such as frequency-dependent dispersion is negligible (as stated in~\citet{Donner202036msp}). Furthermore, the effects of ionosphere are extremely tiny when compared with the effects of IISM. So, the modelling of chromatic red noise inherently covers the modeling of ionosphere. For context, the DM contribution due to ionosphere is generally of the order of 1e-5 pc cm$^{-3}$ (see equation 22 in \cite{lunaska}), which is much smaller than both the SW and IISM effects. So in this study, we neglect the effects of ionosphere to be large enough to require a separate modelling for that effect.

\subsection{Solar Wind Gaussian process}
\label{swgp}
The Solar Wind Gaussian process (SWGP) is a noise model included in the \texttt{enterprise\_extensions}\footnote{https://github.com/nanograv/enterprise\_extensions.git} chromatic noise analysis module that we test here for the first time. It adopts a power spectral density  achieved through a power-law formulation, similar to equation \ref{psdpowerlaw},:
\begin{equation}
\label{psdsw}
    S^{SW}_P(f;A_{SW},\gamma_{SW}) = A_{SW}^2(f/yr^{-1})^{-\gamma_{SW}} yr^3,
\end{equation}
where $S^{SW}_P$ is power-law spectral density for SWGP, $A_{SW}$ is the amplitude of the solar wind at 1/1yr frequency, f the spectral frequency and $\gamma_{SW}$ is the spectral index of secular SW variations.

\begin{figure}[!h]
    \centering
    \includegraphics[width=0.5\textwidth]{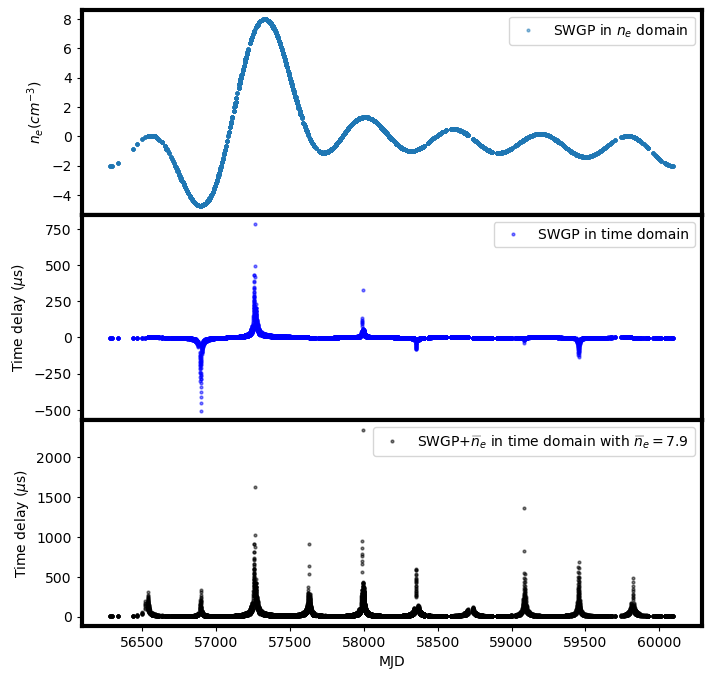}
    \caption{Sample results from one SWGP realisation. In a hypothetical scenario, considering the long timespan of one of the pulsars in our dataset PSR J1022$+$1001. The top panel shows the variations in $n_{\rm e}$ that can be modelled using SWGP perturbed from the mean value. The middle panel shows the corresponding time delays due to SWGP. The bottom panel shows the time delay when the deterministic signal with an $\widebar{n}_{\rm e}$ of 7.9 is added on top of SWGP.}
    \label{fig:swgp_basis_ne}
\end{figure}

The Fourier basis components for SWGP are modulated by the spherical model to account for secular SW variations. Thus, the Fourier components can be expressed as:
\begin{equation}
\label{fourierswgp}
    F^{SW}_l = F^{RN}_l\left({ DM_{\rm{SW,n_e=1}}}\right) \left(\frac{1}{K_D\nu^2}\right),
\end{equation}

where $l=1,\,2,\,\ldots,\,N$. Here, $N$ represents the number of fourier components considered, $F^{RN}$ represents the Fourier components for Red Noise which are formulated according to equation \ref{frs}, $\nu$ is the observing frequency and ${ DM_{\rm{SW,n_e=1}}}$ is the DM due to SW (see eq. \ref{spherical}) when calculated with $n_e =1.0$. This multiplication is needed to effect the impact of the changing line-of-sight impact parameter throughout the year as shown in Figure \ref{fig:swgp_basis_ne}. Using equations \ref{psdsw} and \ref{fourierswgp}, the corresponding time-domain covariance matrix for SWGP can be written as:

\begin{equation}
    \Vec{C}_{\rm{SW}}(t_i,t_j;\theta_s) = \sum_{k,l}^N\ F^{SW}_k(t_i)\Sigma^{SW}_{kl}(\theta_s)F^{SW}_l(t_j).
\end{equation}

Here, $F^{SW}$ represents the Fourier basis functions as detailed in equation \ref{fourierswgp}, $\theta_s = (A_{SW}, \gamma_{SW})$ and $\Sigma^{SW}_{kl}$ is the covariance matrix in frequency domain which is expressed as in equation \ref{covarfreq}. The Fourier components are equally spaced within the frequency range from $1/T$ to $N/T$, with frequency bins truncated at $1/1.5 \;\rm years$. This truncation implies that the SWGP primarily accounts for the lower frequency bins up to the truncation limit, focusing on modeling long-term variations in the solar wind while ignoring shorter time-scale variations. This was implemented into the model due to various reasons. Firstly, the long-term variations in the SW primarily show a cycle of 11 years and at higher frequencies degrade into a red-noise turbulence spectrum. Due to the Gaussian measurement noise present in our data, we generally do not have sensitivity to SW variations at higher frequencies, where they have power below our noise floor. Secondly, since our data are effectively only sensitive to the SW during Solar conjunction (and few weeks before/after), variations faster than $1/2 \;\rm years$ cannot generally be reliably measured. Finally, one important heliospheric component that could be significantly measured in our data at shorter timescales, are Coronal Mass Ejections (CMEs).  In practice, observations affected by CMEs are treated as outliers in long-term pulsar timing studies and are thus excluded from the analysis. Consequently, the effect of shorter time-scale variations on the timing residuals is minimal and does not significantly impact the results.

\subsection{Deterministic signal}
\label{deter}
While previous subsections addressed stochastic noise processes, here we focus on the deterministic noise component of our model. In addition to the secular variations in the solar wind density modeled in \S\ref{swgp}, our model includes a time-constant, average Solar-wind signature quantified by the parameter $\widebar{n}_{\rm e}$, sampled according to the spherical model described by Eq. \ref{spherical}. When combined with the variations quantified by the SWGP term, this fully describes the impact of the solar wind on the pulsar timing observations. Figure~\ref{fig:swgp_basis_ne} demonstrates how this average term and the variable SWGP terms compare, based on the example of PSR~J1022+1001: the top panel shows the time-varying electron density $\Delta n_{\rm e}$ modeled by the SWGP. Since the impact of this parameter on the timing depends on the solar angle (see Equation~\ref{spherical}), the variations in the top panel result in timing delays shown in the middle panel of Figure \ref{fig:swgp_basis_ne}. Since SWGP only models the time-variable component of the solar wind, it can go both positive and negative. But in combination with a constant, $\widebar{n}_{\rm e}$ of 7.9 cm$^{-3}$, the corresponding time delays are depicted in the bottom panel. The equation for the deterministic component of the signal is as follows:

\begin{equation}
    \mathcal{D}(t_i,\nu_i;\widebar{n}_{\rm e}) = \widebar{n}_{\rm e}\frac{\rho(t_i)\  [\mathrm{1 AU}]^2}{r_e(t_i) \sin[\rho(t_i)]}   \frac{1}{ K_D \nu_i^2},
\end{equation}

where $t_i$ is the observing epoch, $\rho(t_i)$, $r_e(t_i)$ and $\nu_i$ represent the pulsar-Sun-observer angle, the distance between the observatory and Sun and observing frequency at $t_i$ respectively. $K_D$ is the Dispersion Constant (see equation \ref{timedel}). Henceforth, note that $n_{\rm e}$ referes to the \textit{varying} electron density at 1 AU whereas, $\widebar{n}_{\rm e}$ refers to the average electron density at 1 AU.

\subsection{Software}

In this subsection, we describe the tools that are used in this study.

\textbf{{\texttt{TEMPO2} and \texttt{libstempo}:}}
\texttt{TEMPO2} (\citealt{Hobbsetal2006a}, \citealt{Hobbsetal2006b}) is a software package utilized for the analysis of pulsar ToAs. \texttt{TEMPO2} accounts for various effects using different parameters. It facilitates fitting for different parameters using the timing model. \texttt{libstempo}\footnote{https://github.com/vallis/libstempo.git} is a Python wrapper of \texttt{TEMPO2}. This package provides the same functionalities as \texttt{TEMPO2}, with the added advantage of seamless integration with other Python-based software. In this work, \texttt{libstempo} has been employed to simulate residuals to evaluate the SW model detailed in \S\ref{simulations}.

\textbf{{\texttt{enterprise} \& \texttt{enterprise\_extensions}}:}
This package has been developed to model the noise processes in the timing residuals based in python. All the noise models that are used in this study are embedded into this software. We use a \texttt{PTMCMCsampler}\footnote{https://github.com/nanograv/PTMCMCSampler.git} to conduct the Markov-chain Monte Carlo sampling. 

\textbf{{\texttt{laforge}:}}
\texttt{laforge}\footnote{https://github.com/nanograv/la\_forge.git} \citep{la-forge} is used to create time-domain realizations from the models defined in \texttt{enterprise} that are parameterised in the frequency domain. In this study we apply this method to various noise models like achromatic red noise, DM noise and SWGP. A detailed description of this package is in Iraci et al (in prep).

\begin{table}[H]
\renewcommand{\arraystretch}{1.7}
    \centering
    \begin{tabular}{|c|c|} \hline 
         \textbf{Parameters}& \textbf{Priors}\\ \hline 
         EFAC (by backend)& $\mathcal{U}(0.1,5.0)$\\ \hline 
          EQUAD (by backend)& $\mathcal{U}(10^{-8},10^{-2})$\\ \hline 
         $\rm log_{10}(A_{DM})$& $\mathcal{U}(-18,-4)$\\ \hline 
         $\gamma_{DM}$& $\mathcal{U}(0,7)$\\ \hline 
         $\rm log_{10}(A_{RN})$& $\mathcal{U}(-20,-8)$\\ \hline 
         $\gamma_{RN}$& $\mathcal{U}(0,7)$\\ \hline 
         $\rm log_{10}(A_{SW})$& $\mathcal{U}(-12,1)$\\ \hline 
         $\gamma_{SW}$& $\mathcal{U}(-6,5)$\\ \hline 
         $\widebar{n}_{\rm e}$& $\mathcal{U}(0,25)$\\ \hline
    \end{tabular}
    \caption{Parameter list with the corresponding priors. $\mathcal{U}$ represents a uniform prior and the ranges are given in the parentheses. EFAC and EQUAD correspond to the WN parameters. $A$ and $\gamma$ for each noise parameter represent the amplitude and spectral index. $\widebar{n}_{\rm e}$ represents the deterministic signal (average electron density) which is considered in addition to the variations offered by SWGP. Note that the priors for amplitude of each of the parameters are in log-uniform space.}
    \label{tab:prior_table}
\end{table}

\section{Simulations}
\label{simulations}
We tested the SWGP implementation described in the \S\ref{swgp} on simulated ToA datasets using \texttt{libstempo}. In particular, we generated noise-free ToAs with SW signal. To introduce statistical errors, we injected reasonable amount of white noise into the simulations which is detailed in each scenario for simulations. Furthermore, to make the simulations more realistic, we injected RN and DM noise with probability distribution that is a power-law in form. The corresponding amplitudes and slopes for RN and DM noise were drawn from probability distributions well established literature (see e.g. \citet{goncharov2021}).

The observational characteristics of these simulations were specifically tailored to those of the LOFAR telescope. This means an acquisition cadence spanning approximately 3 to 5 days, and a radio-frequency coverage from 110 to 190 MHz with 10 subbands. The total temporal baseline of the simulations spans $4$ years. The pulsar on which we based the simulations was PSR J1022+1001; this pulsar has been monitored by all PTA experiments for several decades. Its ecliptic latitude is $-0.06^{\circ}$ making it an ideal source for investigating the SW. Note that we do not consider CMEs in the simulations and other shorter term variations as they are considered as outliers in the PTA datasets as explained in \S\ref{swgp}. Our simulations cover four different test cases, which are detailed in the following subsections.

\subsection{Simulations with yearly-varying $n_{\rm e}$}
\label{yearly}

In this subsection, we describe two different scenarios with various degrees of complexity. In scenario 1, we have a SW signal without any additional noise and in Scenario 2, we attempt to include other kinds of noise like WN, RN and DM noise.

\subsubsection{Scenario 1}

\begin{figure}[!h]
    \centering
    \includegraphics[width=0.47\textwidth,trim = 0mm 3mm 0mm 3mm, clip]{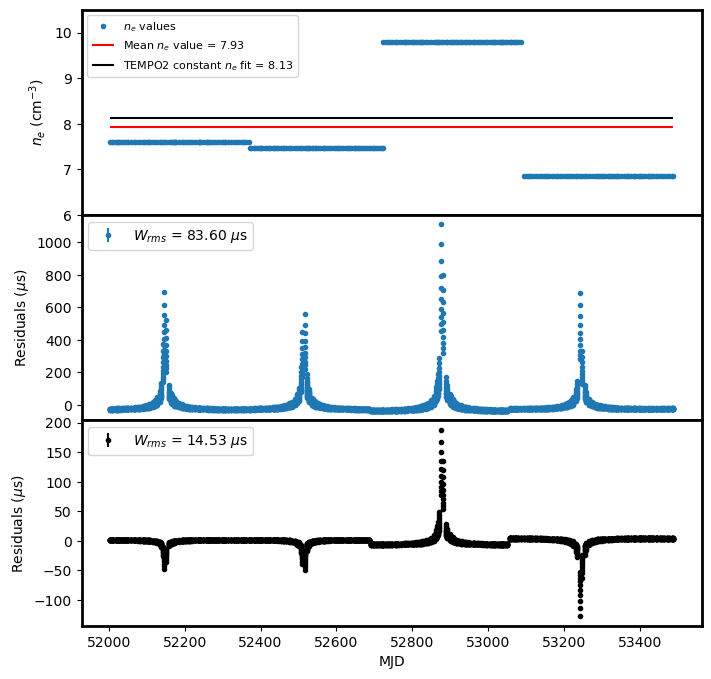}
    \caption{Simulation scenario 1, Only SW. Top Panel: The $n_{\rm e}$ variations simulated (blue) are shown as well as the mean value (red) and the $n_{\rm e}$ retrieved (green) when fit for a single value with \texttt{TEMPO2}. Middle Panel: The blue points represent the residuals  corresponding to the injected values of $n_{\rm e}$. Bottom panel: The green points represent the residuals after a constant $\widebar{n}_{\rm e}$ fit in \texttt{TEMPO2}.}
    \label{fig:step_residuals}
\end{figure}

\begin{figure*}[]
    \centering
    \includegraphics[width=0.8\textwidth,height=0.7\textwidth]{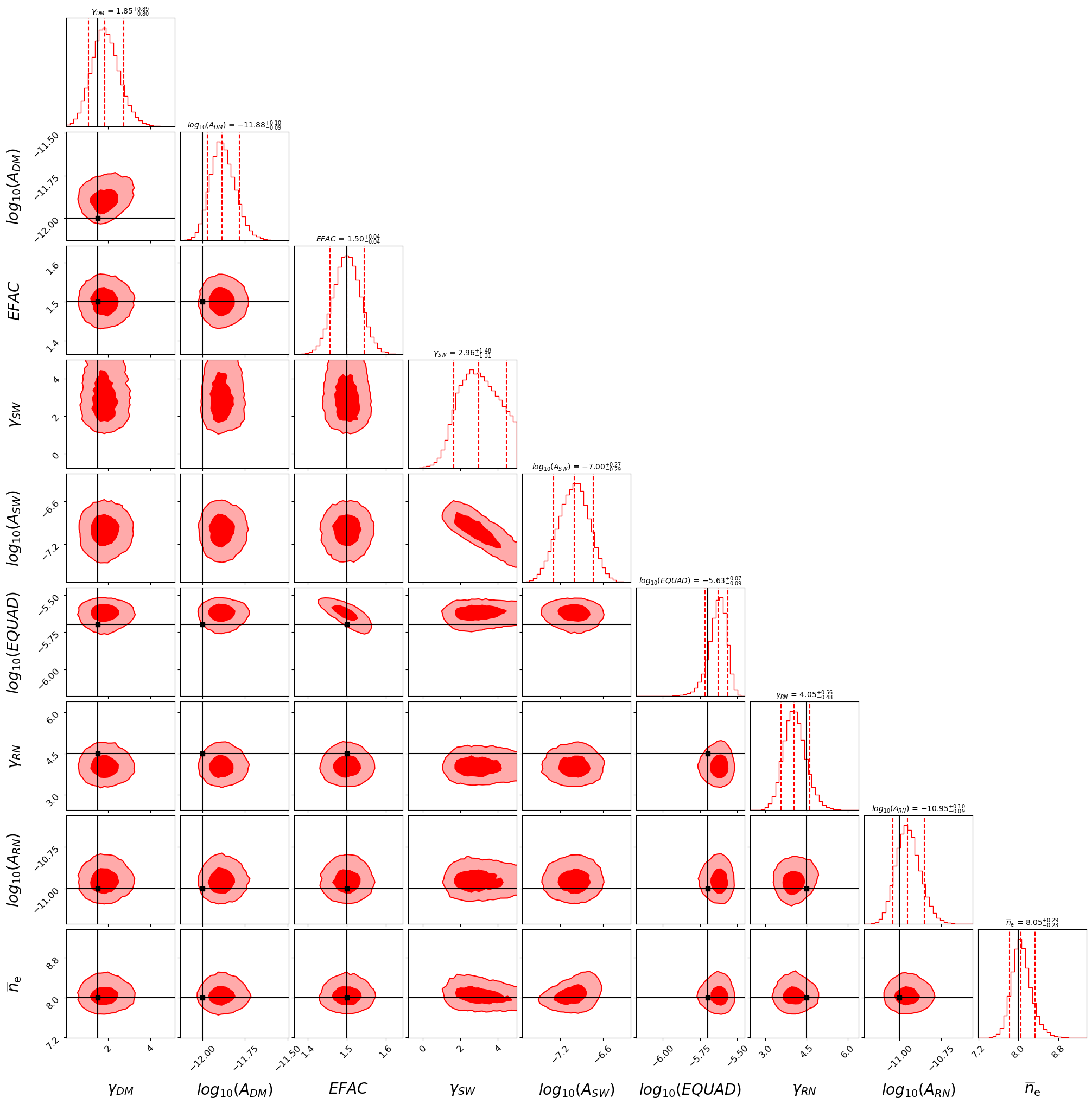}
    \caption{Simulation scenario 2, with noise. Posterior chain plot for yearly varying $n_{\rm e}$ scenario. The black lines represent the truth values injected into the simulations. The black line in $\widebar{n}_{\rm e}$ box is the average value of $\widebar{n}_{\rm e}$ across a period of 4 years. The red lines in the histograms represent the 3$\sigma$ error from the maximum likelihood value.}
    \label{fig:cont_corner}
\end{figure*}

In the first scenario, we simulate the SW amplitude ($n_{\rm e}$) with a constant value per year, with a yearly step function centered on the solar conjuction with the pulsar i.e., $4$ different values as shown in Figure \ref{fig:step_residuals}. In reality the value of $n_{\rm e}$ fluctuates much more rapidly than this but for our first scenario we used this very simple description. We then used Eq. \eqref{spherical} to create a DM timeseries and subsequently the corresponding timing residuals using Eq. \eqref{timedel}. These steps are summarized in Figure \ref{fig:step_residuals}. Following a fit for a single constant value of $\widebar{n}_{\rm e}$ using TEMPO2, we can see that the post-fit residuals (Figure \ref{fig:step_residuals}, bottom panel) clearly demonstrate the necessity of modeling for the variability of the SW.

\subsubsection{Scenario 2}

Next, to improve the noise-free scenario and make the yearly scenario more realistic, we also included three sources of noise, whose parameters are reported in Table \ref{tab:sims_step_noise}.

\begin{table}[h]
    \centering
    \begin{tabular}{|c|c|c|} \hline  
         \textbf{White Noise}&  \textbf{DMGP}& \textbf{Red Noise}\\ \hline 
         \\[-1em]
 $EFAC = 1.5$ 
& $A = 10^{-12}$&$A = 10^{-11}$ \\ \hline 
        \\[-1em]
         $EQUAD = 2 \times 10^{-6}$&    $\gamma$ = \textit{1.5}&  $\gamma$ = \textit{4.5}\\ \hline 
    \end{tabular}
    \caption{Parameters of the  noise components injected in Scenario 2 of the first test case.}
    \label{tab:sims_step_noise}
\end{table}

\begin{figure}[h]
    \includegraphics[width=0.45\textwidth]{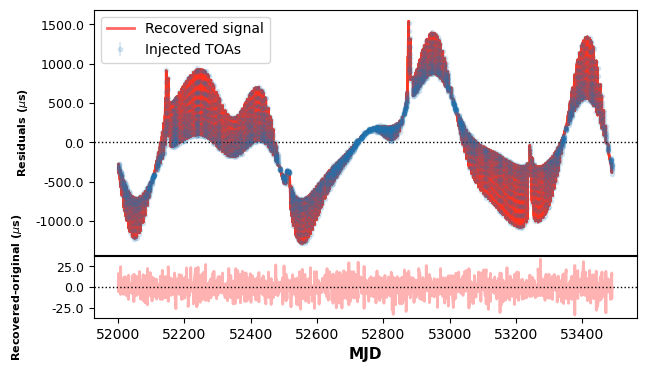}
    \caption{The time domain reconstruction of the inserted simulations with various noise parameters embedded into it. It has the specified noise parameters as in Table \ref{tab:sims_step_noise} along with the SW signal. The blue points are the simulated residuals and the orange lines are the recovery after the noise analysis. }
    \label{fig:step_recons}
\end{figure}

We then re-performed a single pulsar noise analysis using the formulation detailed in \S\ref{model}. We recover the noise parameters and, by using the SWGP approach, we additionally recover the $\widebar{n}_{\rm e}$ value. The resulting posterior plot (Figure \ref{fig:cont_corner}) visually illustrates the fidelity of the parameter recovery, with the black lines within the posterior plot representing the noise values originally injected. With respect to the $\widebar{n}_{\rm e}$ parameter, the black line signifies the average of all injected values of $n_{\rm e}$, as shown in Figure \ref{fig:cont_corner}. This indicates that the SWGP parameters effectively account for an annually-varying $n_{\rm e}$, making the recovered value of $\widebar{n}_{\rm e}$ in the posterior chain an average of injected values.

To check for the accuracy of this posterior parameters, we employed the \texttt{laforge} software package which reconstructs the Gaussian process parameters to a time-domain signal shown in Figure \ref{fig:step_recons}. Note that the degree of success in recovering the DM noise varies with the amplitude of the RN process (Iraci et al. in prep).

\subsection{Simulations with continuously varying $n_{\rm e}$}
\label{contne}
\begin{figure}[h]
    \includegraphics[width=0.45\textwidth]{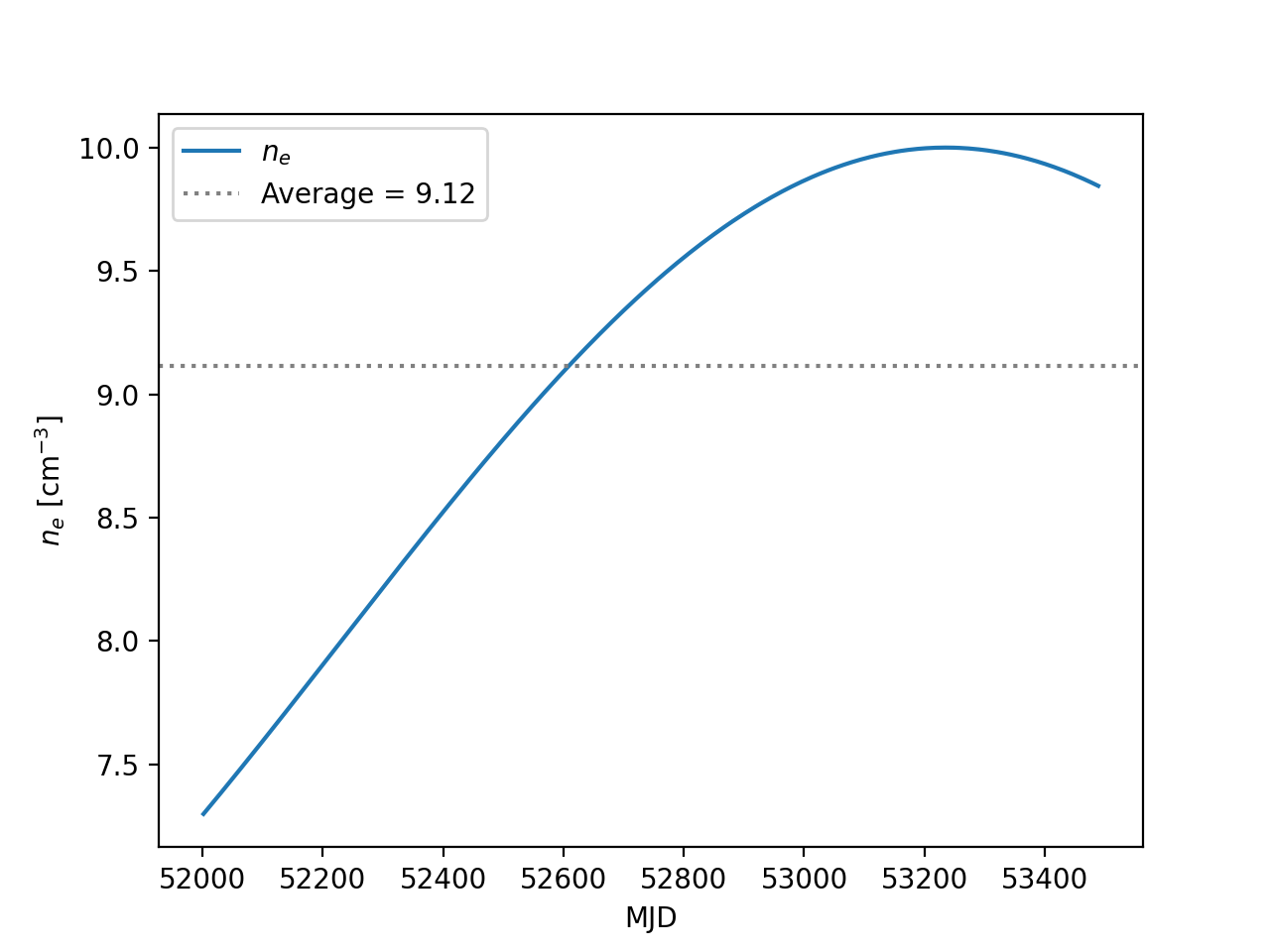}
    \caption{$n_{\rm e}$ plotted as a function of MJD for continuously varying SW scenario.}
    \label{fig:cont_nesw}
\end{figure}

In this case, the $n_{\rm e}$ that we injected in the simulations was modulated on a daily basis for a period of 4 years, following a sine wave pattern with an 11-year periodicity to broadly mimic the solar activity cycle. Subsequently, these $n_{\rm e}$ values were used to construct a DM time series, characterized by  Eq \ref{spherical}. Figure  \ref{fig:cont_nesw} shows the values of $n_{\rm e}$ as a function of MJD. The parameters of the other noise processes that we injected in the simulated ToAs are shown in Table \ref{tab:sims_noise} below. 

\begin{table}[h]
    \centering
    \begin{tabular}{|c|c|c|} \hline  
         \textbf{White Noise}&  \textbf{DMGP}& \textbf{Red Noise}\\ \hline 
         \\[-1em]
 $EFAC = 1.5$ 
& $A = 10^{-13.5}$&$A = 10^{-11.5}$ \\ \hline 
        \\[-1em]
         $EQUAD = 2 \times 10^{-6}$&    $\gamma$ = \textit{1.5}&  $\gamma$ = \textit{4.5}\\ \hline 
    \end{tabular}
    \caption{Parameters of the noise components injected in the simulations that assume a continuously varying $n_{\rm e}$ value. The reduction of amplitudes of DMGP and Red Noise from the previous scenario are deliberately used to make the effects of SW more visible on the residuals in this case (See Figure \ref{fig:contnesw_recons}).}
    \label{tab:sims_noise}
\end{table}

\begin{figure}[h]
    \includegraphics[width=0.5\textwidth]{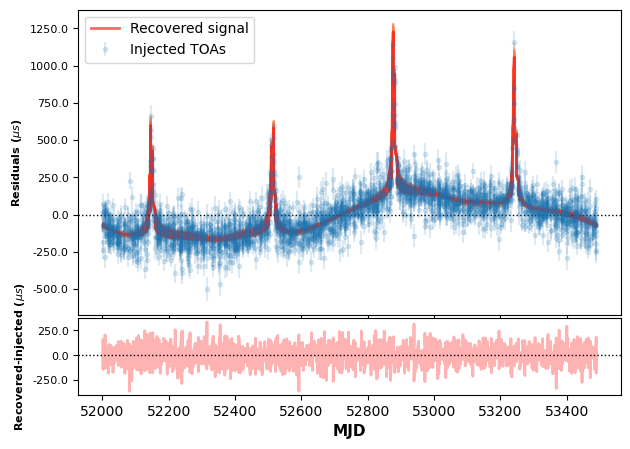}
    \caption{Time domain reconstruction from the continuously varying SW scenario.}
    \label{fig:contnesw_recons}
\end{figure}

Figure \ref{fig:contnesw_recons} shows the time-domain reconstructed signal from the posterior chain that was obtained with a reduced $\chi^2_{\rm red}$ of 1.46.

The foregoing analyses establish the efficacy of employing SWGP for modeling varying $n_{\rm e}$ value across the entire timespan. The fit with \texttt{TEMPO2} becomes dependent upon the number of epochs within a given solar conjunction which inherently biases the fit towards the value associated with a greater number of epochs. In contrast, modeling the variability using SWGP effectively caters to the variability, ensuring robustness in the modeling process. In the next subsection, we demonstrate the usage of SWGP on the two-phase model.

\subsubsection{Extracting the electron densities at 1 AU}
\label{extractne}

The cyclic nature of the Sun's magnetic field, which undergoes reversal approximately every 11 years, induces a correlated fluctuation in the occurrence of substantial solar eruptions, such as solar flares. These eruptions emit bursts of energy and matter into space. In this scenario, our objective is to emulate these natural variations by modulating the $n_{\rm e}$. In our setup, we replicate the 11-year sine wave but with a timespan of 12 years. This deliberate choice was made to encompass an entire solar activity cycle and to meticulously evaluate the ability of our simulations to accurately recover $n_{\rm e}$.

\begin{figure}[h]
    \includegraphics[width=0.5\textwidth]{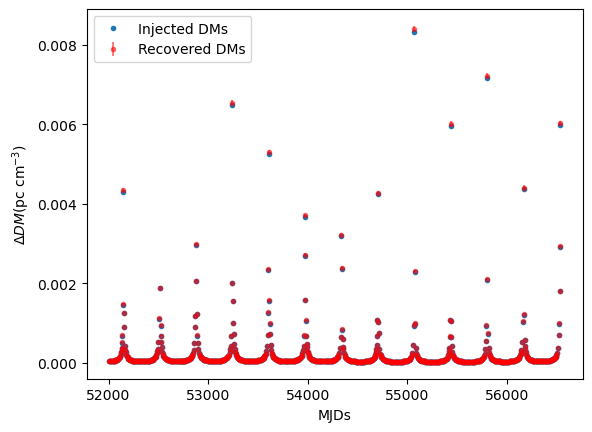}
    \caption{DM timeseries for replicating the solar activity in $n_{\rm e}$. The blue points are injected ones  while the red points with the error bars are recovered from the noise analysis run after modeling using SWGP.}
    \label{fig:dm_t_cont_nesw}
\end{figure}

Figure \ref{fig:dm_t_cont_nesw} illustrates the comparison between the injected DM timeseries and the recovered DMs obtained from posterior chain sampling SWGP. Moreover, we inverted the eq. \ref{spherical} to reconstruct the sine curve of $n_{\rm e}$ injected into the simulations as follows:

\begin{equation}
    n_{\rm e}^{recovered} = \frac{DM_{recovered}}{(1 AU)^2}\frac{|r| sin\rho}{\rho}
    \label{spherical_inv}
\end{equation}

\begin{figure}[h]
    \includegraphics[width=0.5\textwidth]{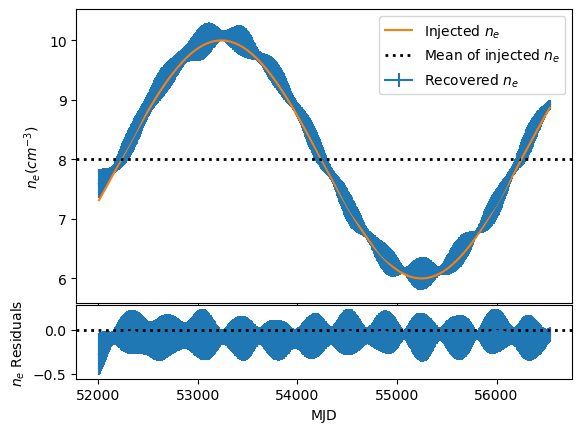}
    \caption{$n_{\rm e}$ recovered from the recovered DM timeseries in Fig \ref{fig:dm_t_cont_nesw}. The orange sine curve is the $n_{\rm e}$ injected into the simulations mimicking the solar activity cycle. The black dotted line represented the average of all injected $n_{\rm e}$.}
    \label{fig:nesw_cont_nesw}
\end{figure}

In Figure \ref{fig:nesw_cont_nesw} we present the injected and recovered values of $n_{\rm e}$. The orange curve represents the injected $n_{\rm e}$ curve, while the blue curve with accompanying error bars illustrates the recovered values of $n_{\rm e}$. Notably, during the solar conjunction of the pulsar, when the LoS of the pulsar is in close proximity to the Sun, the error bars diminish, indicating increased sensitivity to variations in the DMs induced by SW effects. This is also shown in the LOFAR data in Figure \ref{fig:neswts_all}. Conversely, elsewhere, the error bars exhibit an oscillating pattern, with a period of 1 year, as depicted in the bottom panel of Figure \ref{fig:nesw_cont_nesw}.

\subsection{Simulations using the two-phase model}
\label{twophase}
In the third scenario, we reproduced a physical model originally presented by \citealt{You2007}, that is based on the bimodal nature of the solar wind.

\paragraph{On the two phase model:} The SW can be conceptualized as comprising a quasi-static component, which exhibits a bimodal distribution and co-rotates with the Sun, and a transient component, characterized by shorter time scales ranging from hours to days. Our main focus lies on the bimodal co-rotating structure of the solar wind. This structure consists of fast and slow components, each characterized by distinct velocity and density profiles. Specifically, the density profiles are defined by the following equations:

$$n_{\rm e}^{fast} = \bigg[1.155 \times 10^{11} \bigg(\frac{R}{R_{\odot}}\bigg)^{-2} + 32.3 \times 10^{11} \bigg(\frac{R}{R_{\odot}}\bigg)^{-4.39}$$ 
$$+ 3254 \times 10^{11} \bigg(\frac{R}{R_{\odot}}\bigg)^{-16.25}\bigg] m^{-3},$$

$$n_{\rm e}^{slow} = \bigg[2.99 \times 10^{14} \bigg(\frac{R}{R_{\odot}}\bigg)^{-16} + 1.5 \times 10^{14} \bigg(\frac{R}{R_{\odot}}\bigg)^{-6}$$ 
$$+ 4.1 \times 10^{11} \bigg(\bigg(\frac{R}{R_{\odot}}\bigg)^{-2} + 5.74 \bigg(\frac{R}{R_{\odot}}\bigg)^{-2.7}\bigg)\bigg] m^{-3},$$

where $R_{\odot}$ is solar radii and $R$ is the distance from the sun expressed in solar radii (\citealt{guhathakurta}, \citealt{muhlemann}).
To compute the DM, we integrate along the LoS using Eq. \eqref{dmeq}. This model utilizes Carrington rotation maps of the Sun obtained from the Wilcox Solar observatory (WSO) \footnote{http://wso.stanford.edu/} to estimate the LoS. During observation, if the LoS falls within $20^{\circ}$ of the magnetic neutral line, it is classified as slow wind, while other regions are categorized as fast wind. Therefore, both of the phases are taken into account.

\begin{figure}[h]
    \includegraphics[width=0.5\textwidth]{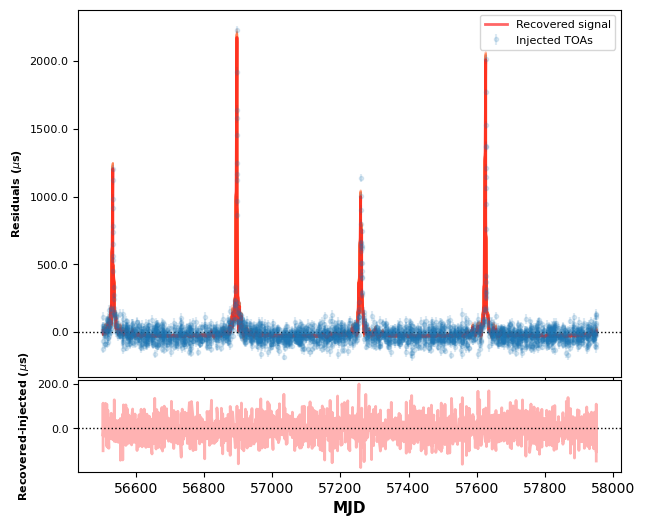}
    \caption{Time domain reconstruction from the simulations when the two-phase model was used. These simulations have WN and the SW noise.}
    \label{fig:you_wn}
\end{figure}

The results of these simulations have been highly useful in our understanding of the implementation of SWGP. Although acknowledging its limitations in fully capturing SW complexities (as highlighted in \citealt{tiburzi2019}), it is worth noting the utility of this model in generating simulations, as it is not based on the spherical model. 

\paragraph{On the implementation of simulations:} We first generate a DM time series using the bimodal structure of SW, assuming to be targeting PSR~J1022+1001 for the 4 years of the simulation timespan. Subsequently, the ToAs were derived using Equation \eqref{timedel}. To introduce statistical uncertainty, we injected WN into the simulations with an EFAC of 1.5 and an EQUAD of $2\times10^{-6}$. Additionally, leveraging both the WN and SWGP components of the model, we generated a posterior chain. The exclusion of other sources of noise was deliberate, aimed at assessing the effectiveness of SWGP in modelling SW independent of a spherical model. The time domain reconstruction plot, illustrated in Figure \ref{fig:you_wn}, presents our findings. Our analysis resulted in a reduced $\chi^2_{\rm red}$ value of 2.2. 

\subsection{Conclusions on simulations}
The simulations demonstrate SWGP's effectiveness. The $\widebar{n}_{\rm e}$ parameter in the posterior chain captures the constant SW signal, while SWGP parameters model the variability of the injected SW perturbed from the $\widebar{n}_{\rm e}$ value. We will apply this model to the LOFAR dataset in the following sections.

\begin{figure*}[!htp]
    \includegraphics[width=1\textwidth,height=0.65\textwidth]{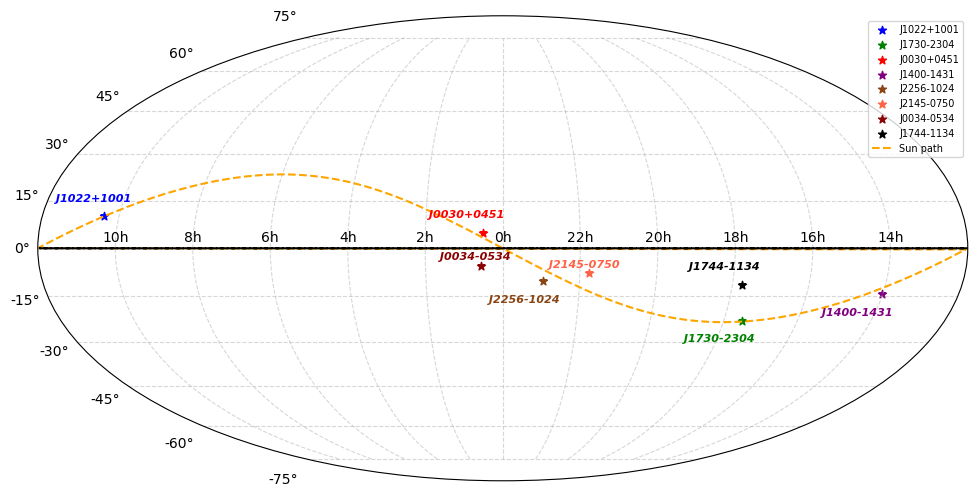}
    \caption{Locations of the pulsar in reference to the Sun in equatorial coordinates. The yellow line represents the path of the Sun and the colored dots represent the locations of each pulsar.}
    \label{fig:skymap}
\end{figure*}

\section{LOFAR dataset}
\label{LOFARdata}

The data used in this study were collected from various pulsar monitoring campaigns spanning approximately 11 years using the LOFAR core telescope \citep{vanhaarlem2013} and single station use of the German LOng Wavelength (GLOW) group\footnote{\url{https://www.glowconsortium.de/index.php/en/}}. The observing bandwidth encompasses a frequency range from 100 to 190 MHz, with a central frequency of about 150 MHz. 
All observations were coherently dedispersed, folded into 10-second sub-integrations and channellized in 360 frequency channels 195 kHz-wide, using the \texttt{DSPSR} software suite \citep{vanstraaten2011}. The integration lengths varied, ranging from 1 to 3 hours for observations conducted with the German stations (DE60X), and from 7 to 20 minutes for those conducted with the LOFAR Core telescope. In the post-processing phase, each observation was cleaned of radio-frequency interference using a modified version of the \texttt{COASTGUARD} software package (see \citet{lazaruscoast}, \href{ https://www2.physik.uni-bielefeld.de/fileadmin/user_upload/radio_astronomy/Publications/Masterarbeit_LarsKuenkel.pdf}{Kuenkel (2017)} \footnote{\url{https://github.com/larskuenkel/iterative_cleaner}}) and corrected for the parallactic angle rotation and projection effects via the \texttt{DREAMBEAM} software package\footnote{\url{https://github.com/2baOrNot2ba}}. After this, each observation is time-averaged, and partially frequency-averaged to reach a fixed number of channels (5, 10 or 20) depending on the signal-to-noise (S/N) of the pulsar using the \texttt{PSRCHIVE} software suite \citep{straatenpsrchive,hotanpsrchive}.

Our analysis includes eight pulsars, that were chosen based on the following criteria:
\begin{itemize}
    \item Millisecond pulsar.
    \item Ecliptic latitude between $-15^{\circ}$ and $15^{\circ}$.
    \item A weekly cadence around the SW conjunction.
    \item More than 4 years of observing time span.
    \item Prominent SW detection in \citet{Tiburzi2021} or \citet{donner2020}.
\end{itemize}

\begin{table*}[!ht] 
    \centering
    \begin{tabular}{ccrrrrrl} 
        \toprule
        Pulsar Name 
        & Time span, $t_{\rm span}$ 
        & \multicolumn{2}{c}{Galactic} 
        & \multicolumn{1}{c}{Period} 
        & \multicolumn{1}{c}{Ecl.~Lat.} 
        & \multicolumn{1}{c}{DM} &Stations \\
        (J2000) 
        & (MJD) (yr)
        & \multicolumn{2}{c}{Coordinates (deg)} 
        & \multicolumn{1}{c}{(ms)} 
        & \multicolumn{1}{c}{(deg)} 
        & \multicolumn{1}{c}{(pc/cm$^3$)} &used\\
        \midrule
        J0030+0451 & 56293$-$60097 (10.4 yr)& 113.1 & $-57.6$ & 4.9 & 1.45 & 4.3330 &LC, DE60X\\ 
        J0034$-$0534 & 56286$-$60251 (10.9 yr)& 111.5 & $-68.1$ & 1.8 & -8.53 & 13.7652 &LC, DE60X\\
        J1022+1001 & 56280$-$60094 (10.4 yr)& 231.8 & 51.1 & 16.5 & $-0.06$ & 10.2530 &LC, DE60X\\ 
        J1400$-$1431 & 57302$-$59807 (\phantom{0}6.9 yr)& 327.0 & 45.1 & 3.1 & $-2.11$ & 4.9322 &LC, DE60X\\ 
        J1730+2304 & 56456$-$59982 (\phantom{0}9.7 yr)& 3.1 & 6.0 & 8.1 & 0.19 & 9.6257 &LC\\ 
        J1744$-$1134 & 56281$-$60095 (10.4 yr)& 14.8 & 9.2 & 4.1 & 11.81 & 3.1385 &LC, DE60X\\ 
        J2145$-$0750 & 56293$-$60095 (10.4 yr)& 47.8 & $-42.1$ & 16.1 & 5.31 & 9.0008 &LC, DE60X\\ 
        J2256$-$1024 & 58286$-$60227 (\phantom{0}5.3 yr)& 59.2 & $-58.3$ & 2.3 & $-3.41$ & 13.7760 &LC, DE60X\\ 
        \bottomrule
    \end{tabular}
    \caption{Table detailing the properties of each pulsar, LC is for LOFAR Core and DE60X means any of the GLOW stations namely DE601, DE602, DE603. DE604, DE605 and DE609.}
    \label{tab:pulsar_table}
\end{table*}

The position of the pulsars with respect to the path of the sun for a whole year is shown in Figure \ref{fig:skymap}. For each pulsar's observational dataset obtained from either the German stations or the LOFAR core, the ToAs were derived as follows. First, the brightest observations from a given observatory-specific dataset were time-averaged together to obtain a frequency-resolved, data-derived template of the pulse intensity as a function of the phase longitude. This was then smoothed using a wavelet-based function to obtain an essentially noise-free, frequency-resolved template using \texttt{psrsmooth} algorithm which is included in \textsc{PSRCHIVE}. A set of frequency-dependent ToAs was then obtained by cross-correlating the noiseless template with the observations themselves. In addition, a Huber-regression based routine (described in \citealt{tiburzi2019}) was used to identify and reject outliers, and obtain a final series of frequency-resolved ToAs, for each observatory and for each of the selected pulsars. 
With these ToA series, we proceeded in applying our SWGP analysis, and we obtained the results outlined in the following section. We note that five of the eight candidates here are a subset of the study shown in \citet{Tiburzi2021}. We also show the detailed properties of each pulsar in Table \ref{tab:pulsar_table}.

\section{Application of SWGP to real data}\label{applicationswgp}

Here we describe the application of SWGP to the LOFAR dataset of \S\ref{LOFARdata}, to account for the SW variability. For this, we need to carry out two main steps.

\paragraph{Optimising the components of the noise model (except for SWGP):} First, we conduct a Bayesian noise model selection on all noise components namely the timing model parameters (also includes a constant $n_e$) using analytical marginalization mentioned in \S\ref{tmm}, WN, DM noise and RN (similar to methods used in \cite{eptawm2}). Note that we consider 30 frequency bins for both RN and DM noise in the initial sampling. Initially, the parameters of the selected noise components were sampled in accordance with the methods outlined in \S\ref{model}. After the analysis of the resulting posterior chains, any noise process determined to be redundant (either due to its insignificance or lack of constraints) was excluded from the final noise run. Notably, only for PSR~J1022$+$1001 we discard the RN component in the final analysis. This is because this pulsar exhibited no significant amount of RN, resulting in an unconstrained posterior. On the contrary, other pulsars demonstrated sensitivity to all the considered noise components. Since all parameters for these pulsars were well-constrained, we retained the choice of 30 frequency bins for RN and DM noise in the final analysis.

\paragraph{Optimization of the SWGP model:} After selecting the most favoured components for the noise model, we optimise the number of frequency bins that we will use to model the SWGP spectrum in the final noise run. The default setting of the SWGP module truncates the SWGP spectrum at a frequency of $1/1.5$~years (as specified in \S \ref{swgp}), however, for some pulsars this results in a negative output value for $n_{\rm e}$, which is non-physical. This implies that the reference cutoff frequency of $1/1.5$~years is not optimal for all pulsars, and we have hence optimised it by testing alternative values. We observe three pulsars for which this change is required: {PSRs} J0034$-$0534, J1400$-$1431 and J2256$-$1024. For the latter two, the optimal frequency cutoff is $1/1$ year, that is motivated by the model's preference for capturing shorter time-scale variations in $n_{\rm e}$, possibly due to the shorter timespan compared to other pulsars in our dataset. Instead, PSR~J0034$-$0534 seems to favour the long time scale variations embedded into the first frequency bin only, $1/t_{span}$. Further details on this pulsar are presented in \S~\ref{J0034}. The details of the frequency bin selection is shown in Table \ref{tab:bin_table}.

\vspace{0.5cm}

\begin{figure}[!h]
    \includegraphics[width=0.5\textwidth,height=0.8\textwidth]{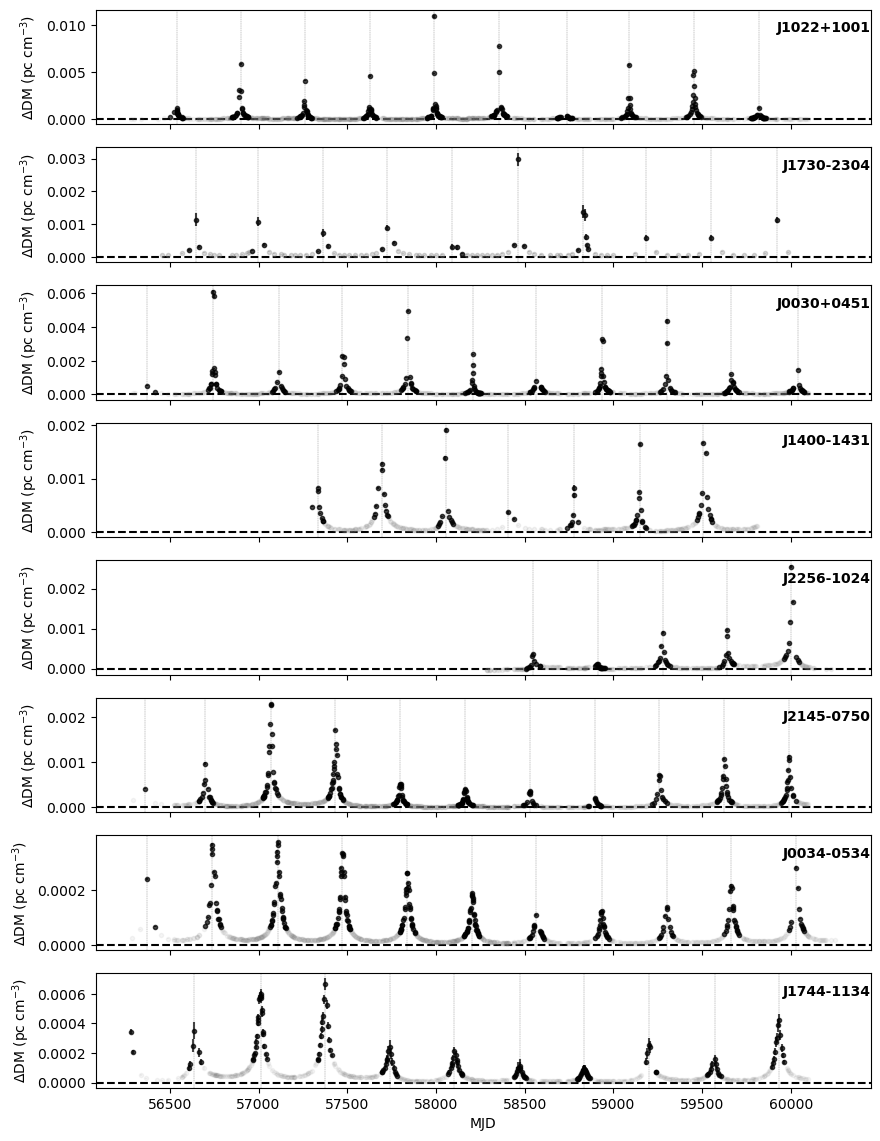}
    \caption{DM time$-$series obtained from the changing SWGP+deterministic $n_{\rm e}$ part of the posterior chain. The black points correspond to epochs with $|solar\_angle| < 45^{\circ}$ from the Sun and the vertical dashed lines indicate Solar conjunction. The order of the pulsars is aligned according to their distance from the Sun in terms ecliptic latitude in ascending order.}
    \label{fig:dmts_all}
\end{figure}

Once the noise analysis is completed, we use the \texttt{laforge} software package to isolate the SW component, namely $A_{SWGP}$, $\gamma_{SWGP}$, and $\widebar{n}_{\rm e}$ and reconstruct the corresponding DM and $n_{\rm e}$ time series (the latter extracted with the method outlined in \S\ref{extractne}). These are shown, respectively, in Figures~\ref{fig:dmts_all} and \ref{fig:neswts_all}.

\begin{table}[h]
\renewcommand{\arraystretch}{1.3}
    \centering
    \begin{tabular}{|c|c|c|} \hline 
         \textbf{Name (J2000)}&  \textbf{Frequency cutoff}& \textbf{No of frequency bins}\\ \hline 
         J0030+0451 &  1/1.5~yr& 6\\ \hline 
         J0034$-$0534 &  1/10yr& 1\\ \hline 
         J1022+1001 &  1/1.5yr& 6\\ \hline 
         J1400$-$1431 &  1/1yr& 6\\ \hline 
         J1730+2304 &  1/1.5yr& 6\\ \hline 
         J1744$-$1134 &  1/1.5yr& 6\\ \hline 
         J2145$-$0750 &  1/1.5yr& 6\\ \hline 
         J2256$-$1024 &  1/1yr& 5\\ \hline
    \end{tabular}
    \caption{Table describing the number of frequency bins used for each pulsar in our dataset.}
    \label{tab:bin_table}
\end{table}

\subsection{Implications of SWGP for pulsar timing}

\begin{figure}[h!]
    \includegraphics[width=0.5\textwidth]{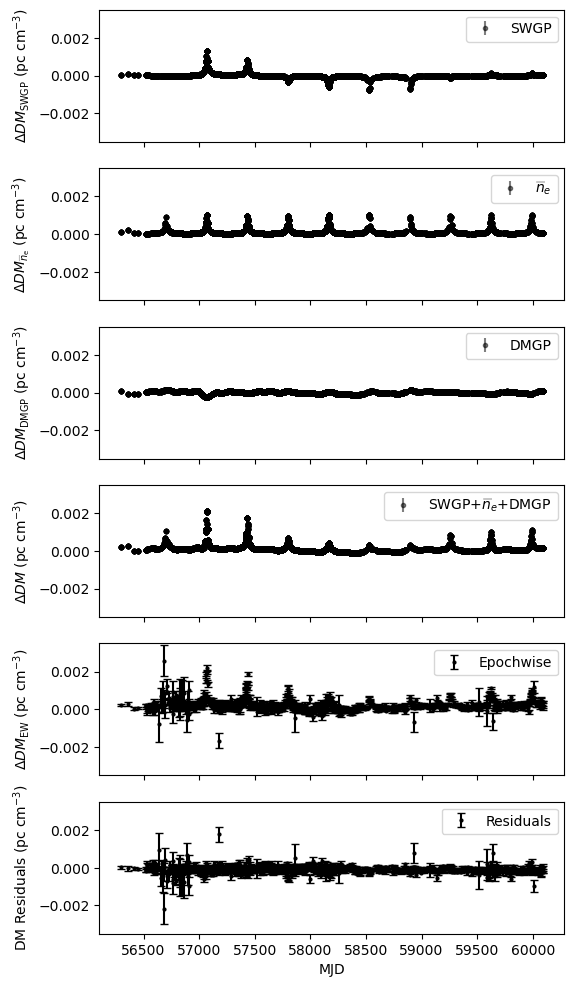}
    \caption{DM timeseries for PSR J2145$-$0750 obtained with different noise elements. The top panel shows the DM reconstruction from the SWGP model. The second panel demonstrates the result from the $\widebar{n}_{\rm e}$ part of the posterior chain. The third panel shows the resultant signal due to DMGP. Assuming these three are the main components that contribute to the DM timeseries, the fourth panel combines these signals. The fifth panel is the DM timeseries using \textit{Epoch-Wise} method. And the bottom panel shows the residuals in DM between the \textit{Epoch-Wise} and the combined signal from SWGP, $\widebar{n}_{\rm e}$ and DMGP. }
    \label{fig:residual_2145}
\end{figure}

In this section, we attempt to deconvolve each noise parameter which contributes to the DM parameter in the DM timeseries for PSR~J2145$-$0750. In Figure \ref{fig:residual_2145}, the top panel consists of the DMs that are modelled using SWGP, the second panel consists of DMs due to the constant average $\widebar{n}_{\rm e}$ from the posterior and the third panel comprises of the DM timeseries due to the DM noise. We attempt to compare the combined DM timeseries from SWGP, $\widebar{n}_{\rm e}$ and DMGP with the DM timeseries obtained with the \textit{Epoch-Wise}\footnote{This method is so-called as it obtains a DM timeseries by fitting the frequency-resolved timing residuals $r_{\nu,i}$ of each observation $i$ as: $$r_{\nu,i} = \frac{DM_i}{\nu^2} + C_i$$ where $C_i$ is an offset that contains any possible achromatic delay introduced by errors in the timing model, and any unaccounted red noise.} method. 

\begin{figure}[h]
    \includegraphics[width=0.5\textwidth]{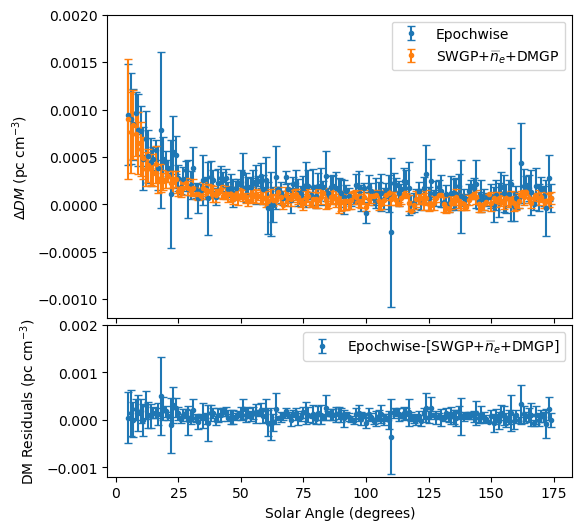}
    \caption{Demonstration of SWGP model in comparison with the \textit{Epoch-wise} method as a function of solar angle on PSR J2145$-$0750. Here, we consider the binned average over 1$^{\circ}$ of solar angle. This elucidates the working of SWGP particularly for epochs less than 30$^{\circ}$ from the Sun.}
    \label{fig:sangle_residual}
\end{figure}

This method is outlined in \citet{tiburzi2019} or \citet{donner2020} to obtain and study its DM time series without the need of additional noise analysis. Note that in the DM timeseries shown in the fifth panel is the \textit{Epoch-Wise} method after correcting for the slope in DM embedded into the timing model. The bottom panel of Figure \ref{fig:residual_2145} shows the DM residuals from the \textit{Epoch-Wise} method. Further demonstration of the SWGP model can be seen in Figure \ref{fig:sangle_residual} where we plot the DM values, using both methods, and their residuals as a function of solar angle. This highlights the SWGP model's capability to accurately characterize the evolving effects of solar wind on pulsar timing residuals. Therefore it appears that SWGP offers significant promise for modeling solar wind in future PTA experiments, particularly with the full integration of LOFAR and IPTA data.

\begin{figure}[!h]
    \includegraphics[width=0.5\textwidth,height=0.4\textwidth]{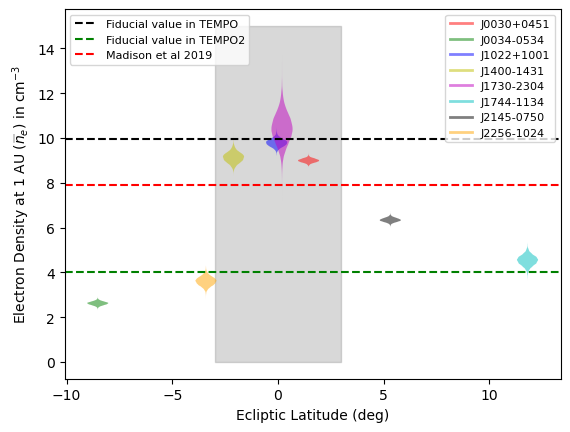}
    \caption{The average electron density, $\widebar{n}_{\rm e}$, as a function of ecliptic latitude after accounting for time variability. Each violin corresponds to a different pulsar with the legends described in the figure. The gray shaded region corresponds to the slow wind region considered in this study (ranging from $-3^{\circ}$ to $3^{\circ}$).}
    \label{fig:nesw_all}
\end{figure}

\subsection{Average Electron density values}\label{averagene}

As we mentioned in the previous sections, traditional pulsar timing analyses use a time-fixed spherical model to account for the SW effects with fixed values of electron density at 1 AU (4 and 9.9681 cm$^{-3}$ for, respectively, \texttt{tempo2} and \texttt{tempo}, and 7.9 cm$^{-3}$ according to \citealt{madison2019}). To have a comparison with this standard approach, we report the average electron density values also denoted as $\widebar{n}_{\rm e}$ for each pulsar, sorted by ecliptic latitude, in Figure \ref{fig:nesw_all}. This plot clearly shows that different $\widebar{n}_{\rm e}$ values affect different pulsars; and that this has a clear dependence on the ecliptic latitude of the pulsar.

Therefore, not only the choice of a time-independent model, but also of a uniform value of $n_{\rm e}$ for pulsars at all ecliptic latitudes in pulsar timing analysis is sub-optimal. A better model describing the SW effects in pulsar timing data needs to account for an ecliptic-latitude dependency in addition to any temporal variations. Figure \ref{fig:nesw_all} also suggests that the pulsars that populate the low ecliptic latitude regions (the grey-shaded area in the plot, highlighting ecliptic latitudes between $-3$ and 3$^{\circ}$) seem to show a systematically higher (and more uniform among different sources) $\widebar{n}_{\rm e}$ value with respect to other pulsars. This might be due to a longer exposure to the slow (and denser) phase of the SW during the solar approach. 
It is important to note that the boundary indicated by the gray-shaded region in Figure \ref{fig:nesw_all} does not represent a strict demarcation for the slow wind. In reality, this boundary is determined by the Heliographic Current Sheet (HCS), which does not coincide with the solar equator. During solar maximum, the HCS becomes highly warped \textit{(Ballerina's skirt shape)} and inclined relative to the ecliptic plane, resulting in a more complex and oblique solar wind stream, in contrast to the distinct bimodal structure observed during solar minimum \citep{poletto2013}. Consequently, intermediate scenarios are likely, wherein pulsar's LoS may pass through different wind streams depending on the tilt of the HCS. Furthermore, this variability can differ between solar activity cycles.

\begin{figure}[!h]
    \includegraphics[width=0.5\textwidth,height=0.8\textwidth]{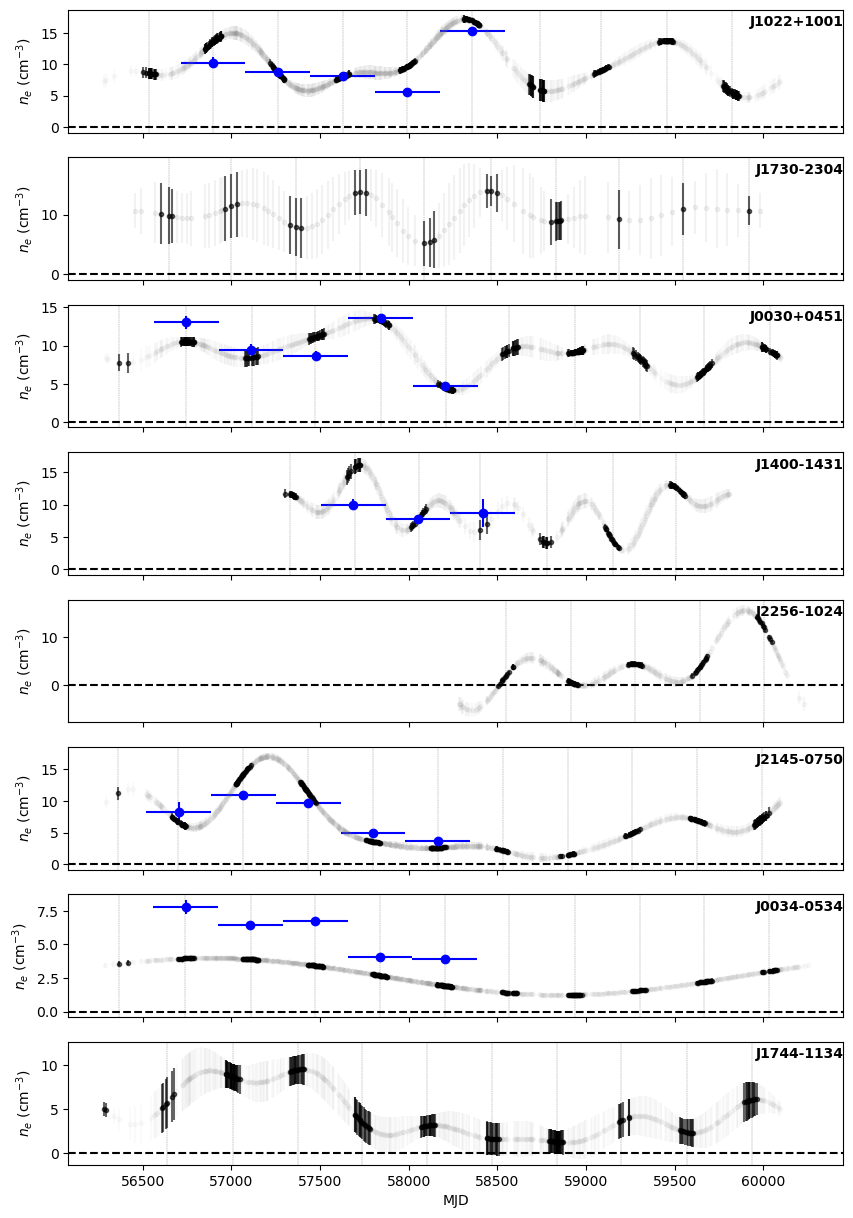}
    \caption{$n_{\rm e}$ extracted from the DM time series shown in Figure~\ref{fig:dmts_all}. The black points represent the epochs that have $|solar\_angle|< 45^{\circ}$. The blue points are obtained from the results presented in \citet{Tiburzi2021}. The order of the pulsars is aligned according to their distance from the Sun in terms ecliptic latitude in ascending order. The long term solar cycle is evident in pulsars more than 3$^\circ$ away from the Sun in terms of ecliptic latitude i.e., the bottom 4 panels.}
    \label{fig:neswts_all}
\end{figure}

\begin{figure*}[!hb]
    \centering
    \includegraphics[width=0.9\textwidth,height=0.80\textwidth]{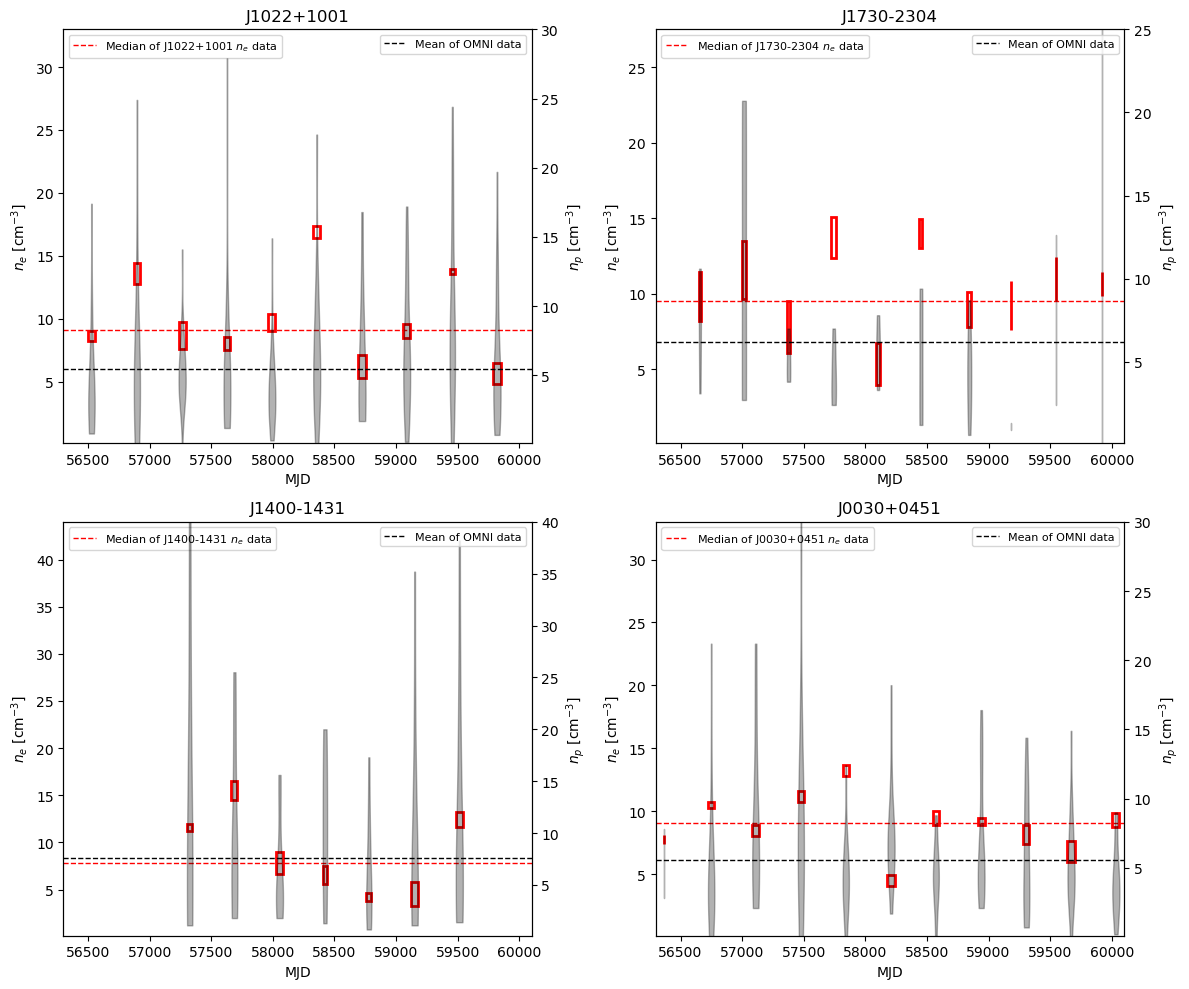}
    \caption{Comparison with In-Situ Observations: The figure shows the variations of $n_{\rm e}$ derived from pulsar observations (red boxes) using the models described in this paper. Only observations where the angular separation between the pulsar and the Sun is less than 30 degrees, as seen from Earth, are included. The red horizontal line represents the median of the pulsar-derived electron density values. The black violin plots indicate the variations in proton density ($n_{\rm P}$) from the OMNI data, measured within a $\pm 2$ hours window around each pulsar observation. The horizontal black dashed line denotes the mean proton density from the OMNI data corresponding to the pulsar observation times. Since the space probes measure proton densities and the pulsars measure electron densities, direct comparison between the two is non-trivial. The four panels represent different pulsars: J1022+1001, J1730-2304, J1400-1431, and J0030+0451, each showing unique density variations over time.}
    \label{fig:neswts_sp}
\end{figure*}

\subsection{Temporal variability in $n_{\rm e}$}

From the $n_{\rm e}$ time series in Figure \ref{fig:neswts_all}, some pulsars such as PSRs~J2145$-$0750 and J1744$-$1134 exhibit a clear correlation with the solar activity cycle, with a peak $n_{\rm e}$ value during the Solar maximum (around MJD 56750, April 2014) and minima during the solar minimum (around MJD 58820, December 2019), in what seems to follow a sine wave with an 11-year period. Other pulsars, on the other hand, like PSRs~J1022+1001 and J0030+0451, do not display any notable 11 year periodicity. In general we can note that pulsars with low ecliptic latitudes seem to have relatively constant $n_{\rm e}$, while sources with medium-to-high ecliptic latitudes tend to show a variability correlated with the Solar cycle.

Physically, such behaviour could be induced by how the two SW phases interact during the Solar cycle, as shown by the Ulysses mission~\citep{mccomas1998}. The Ulysses results showed that during the Solar minimum the SW has a distinct bimodal distribution with approximately well-defined boundaries, with the slow wind being relatively constrained around the neutral magnetic field line and the fast wind elsewhere. However, during the maximum these two phases tend to mix at intermediate heliographic latitudes. We stress that heliographic and ecliptic coordinates are two different reference frames, and that the time-dependent neutral magnetic field line is \textit{not} aligned with the ecliptic. However, the LoS of pulsars with low ecliptic latitudes should mostly pass through the slow wind, especially during the Solar approach and egress\footnote{While at the Solar conjunction any pulsar that is not occulted by the Sun will hover above or below the Solar disk, therefore the LoS at that point will be likely affected mainly by the fast wind}. 

\begin{figure*}[!hb]
    \includegraphics[width=\textwidth,height=0.25\textwidth]{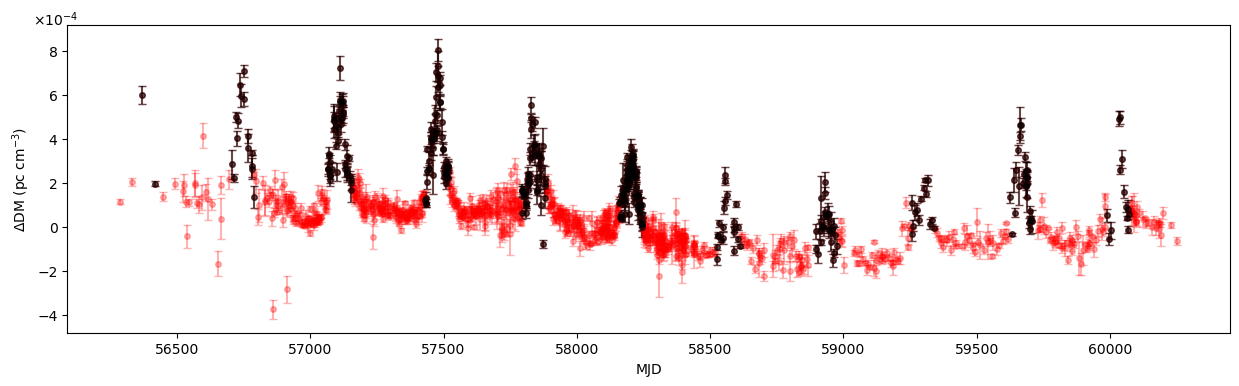}
    \caption{Time series of DM variations for PSR~J0034$-$0534. The black points are the epochs at which the pulsar is closest to the Sun with a solar angle less than $45^{\circ}$. The red points are further away from the Sun}
    \label{fig:dmts_J0034}
\end{figure*}

This is possibly the reason pulsars at low ecliptic latitudes maintain a more stable, and high $n_{\rm e}$ value (due to the presence of a dense SW) throughout the Solar cycle. Moreover, pulsars with intermediate-to-high ecliptic latitudes would be sensitive to the $n_{\rm e}$ variability dictated by the solar cycle in this framework, as we expect the LoS to be passing through the fast wind during the periods of solar minima (causing the $n_{\rm e}$ lows that we observe in Figure~\ref{fig:neswts_all}), and during the solar maxima, the LoS possibly passes through a mixture of fast and slow wind (hence denser than the fast wind alone) causing the highs in $n_{\rm e}$ shown in Figure~\ref{fig:neswts_all}).

\subsubsection{Comparison with the OMNI data}
We compare the $n_{\rm e}$ time series of a subset of pulsars with the proton density at 1~AU\footnote{Due to the presence of He$^{+}$ ions, the proton density is slightly lower than the electron density, but there is no clear consensus on the conversion factor between the two.} provided by the OMNI database \citep{omni}, as obtained with the \textit{in-situ} spacecraft Solar Wind Electron Proton Alpha Monitor (SWEPAM) for the Advanced Composition Explorer (\texttt{ACE}, see \citealt{mccomas1998}) and \texttt{DISCOVR}. In Figure \ref{fig:neswts_sp} we report the results of the comparison, for which we only consider those pulsars which satisfy the condition of $|ELAT|<3^{\circ}$ since the spacecraft cover the ecliptic region in particular. The black violins represent the probe measurements, with the spread of each violin indicating their range of values in $\pm2\ hours$ from the time of each pulsar observation, while the red boxes represent the variations in $n_{\rm e}$ values of each pulsar during the Solar approach. Pulsar observations provide estimates of the integrated free-electron column density, which are subsequently modeled (as described in \S\ref{swgp}) to derive a value of $n_{\rm e}$ at 1 AU. In contrast, the OMNI database provides spot measurements of the free proton density at 1 AU. This fundamental difference accounts for many of the discrepancies between the two datasets, highlighting that the modeling efforts applied to the pulsar data are overly simplistic. While such models may be useful for pulsar-timing experiments \citep{tiburzi2019}, Figure \ref{fig:neswts_sp} demonstrates their inadequacy for space-weather studies in their current form.

The differences between the pulsar-derived electron densities and the in-situ spacecraft measurements of proton densities indicate that our pulsar modeling efforts (which adhere to industry standards used in pulsar timing experiments, e.g., \citet{eptadr2}) are insufficient for accurately modeling space weather. However, pulsar measurements of line-of-sight integrals through the solar wind offer a unique, high-precision contribution to space-weather studies by providing meaningful additional constraints to any solar-wind models across multiple lines of sight, at various ecliptic latitudes, and across a range of ecliptic longitudes. Incorporating such measurements into space-weather models is bound to be complex and challenging, but it holds significant promise.

\begin{figure}[!h]    
\includegraphics[width=0.5\textwidth]{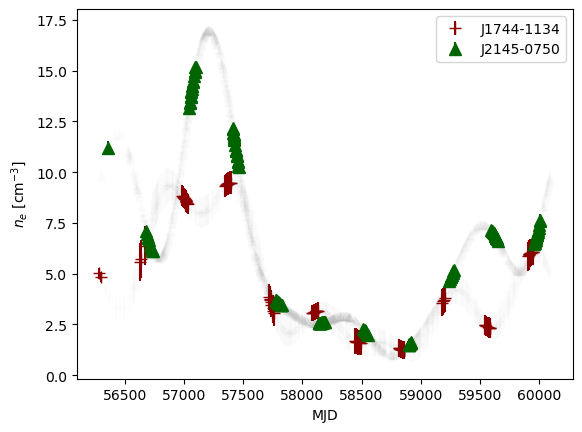}
    \caption{$n_{\rm e}$ obtained from pulsars away from the ecliptic. We discarded PSRs~J0034$-$0534 and J2256$-$1024 from this plot as both of them were categorised special cases.}
    \label{fig:neswothereclat}
\end{figure}

\subsubsection{Sensitivity to the Solar activity cycle}

In Figure~\ref{fig:neswothereclat} we isolated the $n_{\rm e}$ of the pulsars with higher ecliptic latitudes, particularly PSRs~J2145$-$0750 and J1744$-$1134. These pulsars show clear sensitivity to the activity cycle with consistently low values for $n_{\rm e}$ during Solar minima (likely because the LoS is dominated by the fast wind at those times). In contrast, during Solar maxima the LoS of these pulsars may experience various degrees of mixing of the fast and slow winds, leading to different but generally higher values for $n_{\rm e}$.

\subsection{Two special cases: PSRs~J0034$-$0534 and J2256$-$1024}\label{J0034}
PSR J0034$-$0534 is observed by the LOFAR Core and the GLOW stations as part of various pulsar monitoring projects. This pulsar has the highest DM precision among the MSPs in the LOFAR sample (see \citealt{donner2020}), with a median DM uncertainty of $\sim 2.64\times10^{-5}$~pc cm$^{-3}$. The DM timeseries of this pulsar is shown in Figure \ref{fig:dmts_J0034} using the \textit{Epoch-wise} method.

The DM precision is such that the solar wind signal remains noticeable in the data even beyond a solar angle of $45^{\circ}$ and that our measurements have significant sensitivity to the asymmetry in the DM values around the solar conjunction (also pointed out in \citet{tiburzi2019}). Due to this, attempts to incorporate higher-order variations in $n_{\rm e}$ for this pulsar resulted in unphysical models, particularly during anti-solar conjunctions. After verification that these issues were unaffected by the modelling of the other noise processes, we therefore concluded that the best way forward for this pulsar, was to constrain the SWGP model to only describe the long-term variability associated with the solar cycle and to leave the higher-order variations in the solar wind to be modelled by DMGP. Ideally a more complex solar wind model would be used, which can adequately describe the non-spherical nature of the solar wind, but such development was beyond the scope of this paper. Hence, we limit the cutoff frequency to  $1/(10~\textrm{yr})$, which corresponds to $1/t_{\rm span}$. One consequence of this simplistic SWGP model for this pulsar, is that our $n_{\rm e}$ values are systematically offset from those reported by \citet{Tiburzi2021}, as this offset is absorbed in the model for the IISM contribution. Specifically, we posit that the analysis by \citet{Tiburzi2021} erroneously ascribed significant interstellar dispersion variations as arising from the solar wind, while our analysis avoids such leakage.

PSR~J2256$-$1024 also exhibits a complicated scenario. In particular, it has the shortest timespan of our sample (see Figure~\ref{fig:dmts_all}), and the DM reconstruction via \texttt{laforge} shows a number of epochs with negative values, which is nonphysical. Unlike PSR~J0034$-$0534, we suspect this might be due to correlation between the variability in the SW electron density and variations in the interstellar DM. 
We observe that whenever the DM noise is absorbing some SW noise, particularly at the Solar conjunction, SWGP attempts to compensate for that absorption with negative peaks. This covariance with DMGP is an inherent flaw of this model that will have to be taken into account in future analysis.

\section{Application of SWGP as a common signal on real data}
\label{commonsec}
The previous section showed the occurrence of a bimodality in the $n_{\rm e}$ behavior for the pulsars in our sample:
\begin{itemize}
    \item[\checkmark] Pulsars with low ecliptic latitude (between $-3$ and 3 degrees)
    \begin{itemize}
        \item High values of mean $n_{\rm e}$
        \item No clear temporal evolution in $n_{\rm e}$
    \end{itemize}
    \item[\checkmark] Pulsars with medium-to-high ecliptic latitude (from $-20$ to $-3$ degrees and from 3 to 20 degrees)
    \begin{itemize}
        \item Low values of mean $n_{\rm e}$
        \item The temporal pattern of $n_{\rm e}$ correlates with the Solar cycle.
    \end{itemize}
\end{itemize}

This shows that we cannot consider the SW as a common signal among all the pulsars in the sample, however, it seems that it might be regarded as such within these two groups separately. Therefore, we repeat our analysis by configuring the \texttt{enterprise} model to apply the SWGP noise component in common among the pulsars of the first group (encompassing PSRs~J0030+0451, J1022+1001, J1400$-$1431, and J1730$-$2304), and among the pulsars of the second group (all the other sources minus PSR~J0034$-$0534), while maintaining the original settings established in the single-pulsar noise analysis for the other noise elements.

\subsection{Common signal for pulsars with low ecliptic latitudes}
\label{slowcommonsec}

\begin{figure}[!h]
    \includegraphics[width=0.5\textwidth, trim =0mm 2mm 0mm 2mm, clip]{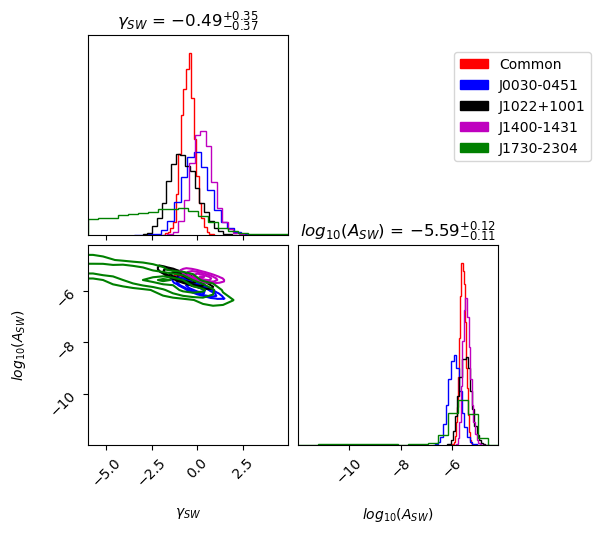}
    \caption{Posterior distributions for the spectral parameters from the SWGP analysis, modelling time variability as a common signal (red) between all pulsars within three degrees of the ecliptic plane; or separately for each pulsar individually (blue, black, purple and green). The titles shown depict the median values for the common noise part corresponding to the red histogram.
    }
    \label{fig:slowcommon}
\end{figure}

Figure \ref{fig:slowcommon} illustrates the posterior distribution of the time-varying component for pulsars with an ecliptic latitude between $-3^{\circ}$ and $3^{\circ}$. 
The red histogram in the corner plot depicts the parameters corresponding to the common signal. The same parameters derived individually for each pulsar (as obtained from the analysis in the previous section and represented by different colors) 
agree well with the outcomes derived from the common noise model. Additionally, the common noise exhibits narrower 
constraint compared to the ones obtained for the individual pulsars, hence supporting its more optimal performance. 
PSR~J1730$-$2304 (depicted in green in Figure~\ref{fig:slowcommon}) displays the broadest posterior distribution, likely due to insufficient observing cadence during the solar conjunctions.

\subsection{Common signal for pulsars with medium-to-high ecliptic latitudes}

\begin{figure}[h]
    \includegraphics[width=0.5\textwidth]{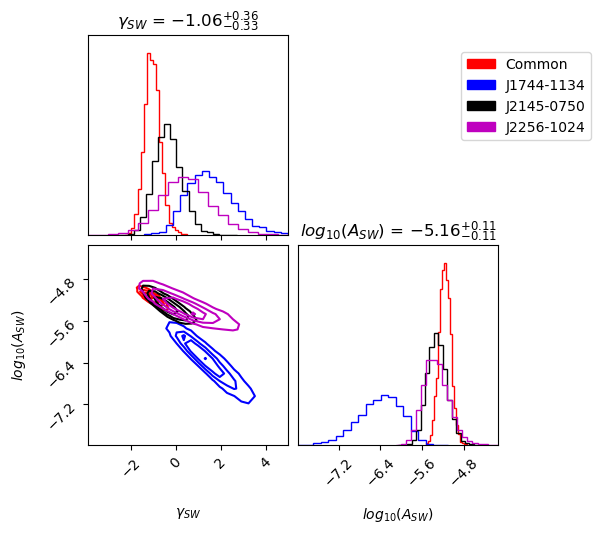}
    \caption{Posterior distributions for the spectral parameters from the SWGP analysis, modelling time variability as a common signal (red) between all pulsars beyond three degrees from the ecliptic plane; or separately for each pulsar individually (blue, black, purple). The titles shown depict the median values for the common noise part corresponding to red histogram.
    }
    \label{fig:fastcommon}
\end{figure}

Figure \ref{fig:fastcommon} displays the posterior distributions for the pulsars in our sample situated the furthest away from the ecliptic region. Once again, the red histogram represents the posteriors for the common SW signal. These distributions indicate that the amplitude of SWGP here is higher when compared to the previous case described in \S\ref{slowcommonsec}, possibly 
suggesting a greater variability. However, there are clear discrepancies between the parameters of the common signal and the ones derived for the individual pulsars. 

In particular, PSR J1744$-$1134 deviates the most from both the other pulsars and the common noise parameters, indicating pulsar-specific variability in amplitude and slope. This deviation can be attributed to PSR J1744$-$1134's ecliptic latitude of $11.81^{\circ}$, i.e. the farthest from the Sun among the pulsars in our study.

PSRs~J2145$-$0750 and J2256$-$1024 are more consistent with the parameters of the common signal within the uncertainty ranges, although not as prominently as in the previous scenario. This experiment underlines the need for caution when considering a common SW signal for pulsars located away from the ecliptic, as their variability seems to imply the necessity for individual treatment. 

\section{Conclusions}
\label{conclusions}

In this study we applied and tested \textit{SWGP}, a Bayesian, Gaussian-process approach to model the time-variable solar wind signal in pulsar-timing data, which approximates the power spectrum of its temporal density variations with a power-law; and its spatial electron-density distribution as spherical. The resulting data products from the application of SWGP are the posterior distribution of both the amplitude and spectral index of the power spectrum, and from these we reconstructed the time series of both the corresponding amplitudes of the spherical SW model at 1~AU, $n_{\rm e}$, and of the corresponding DM time series, using the \texttt{laforge} software package. 

First, we validated the method against pulsar-timing simulations with a progressively increasing level of complexity. The model performed well against all the simulations, including the ones that do not rely on a spherical distribution of free electrons for the SW (see \S\ref{twophase}). We then applied the SWGP module alongside the other commonly used noise components in pulsar-timing noise analysis to study the SW trends in a sample of millisecond pulsars observed for about 10 years with the low-frequency interferometer LOFAR. The analysed observations are particularly sensitive to plasma-related propagation effects. Our results highlight important implications for pulsar timing and related noise analyses, that can be summarised as follows:

\begin{itemize}
    \item[\checkmark] The $n_{\rm e}$ parameter affecting pulsar observations is both dependent on the time of the observation and the ecliptic latitude of the pulsars concerned, in contrast to the assumptions typically used in pulsar timing analyses. This also lays emphasis on the viability of spherical model to account for the effects of SW which does not fully explain the intricacies of this noise process.
    \item[\checkmark] Pulsars with low ecliptic latitude (within three degrees of the plane) have systematically higher values of $\widebar{n}_{\rm e}$ than pulsars with medium-to-high ecliptic latitudes (between three and 20 degrees from the plane) 
    \item[\checkmark] The temporal evolution of inferred $n_{\rm e}$ is different for pulsars at low ecliptic latitudes and for pulsars at medium-to-high ecliptic latitudes. In particular:
    \begin{itemize}
        \item Pulsars with low ecliptic latitude do not show clear $n_{\rm e}$ temporal patterns
        \item Pulsars with medium-to-high ecliptic latitude show a quite distinct temporal pattern in $n_{\rm e}$, that correlates with the Solar cycle, with peaks of $n_{\rm e}$ reached during the Solar maximum, and dips of $n_{\rm e}$ reached during the minimum.
    \end{itemize}
    \end{itemize}

We also confirmed this difference in behavior by using SWGP as a common noise model for pulsars with low ecliptic latitude and medium-to-high ecliptic latitude. This exercise showed that while the posteriors of low ecliptic sources tended to agree with the ones of the common signal, for the second group there was less consistency between the common signal and the individual time variability posteriors. 

The physical reasons of these features probably lies in the bimodal SW nature, that materializes in a slow-wind phase and a fast-wind phase, and in their interaction during the Solar cycle. 
While the verification of this hypothesis is beyond the scope of this article, it may be pursued by evaluating the distribution of slow and fast solar wind along the LoS for each pulsar observation, e.g.~based on white-light 
images of the Solar environment or Solar magnetograms, or through the comparison of the LoS positions within the 3D reconstructions of the SW that can be offered by space-weather software such as EUHFORIA or 3D IPS tomography \citep{tjc23,smt23}. In turn, pulsar observations prove to be optimal bench tests for such software, and might be able to contribute to their functionalities in the near future.

\section*{Acknowledgements}
SCS acknowledges the support of a College of Science and Engineering University of Galway Postgraduate Scholarship in supporting this work.
ACh and GMS acknowledge financial support provided under the European Union’s H2020 ERC Consolidator Grant “Binary Massive Black Hole Astrophysics” (B Massive, Grant Agreement: 818691).
JPWV acknowledges support from NSF AccelNet award No.~2114721. JSH acknowledges support from NSF CAREER award No.~2339728 and NSF Physic Frontiers Center award No.~2020265. 
FI was supported by the University of Cagliari, Italy and the Istituto Nazionale di Astrofisica (INAF). Pulsar research at Jodrell Bank Centre for Astrophysics is supported by an STFC Consolidated Grant (ST/T000414/1; ST/X001229/1). This paper is based on data obtained with the German stations of the International LOFAR Telescope (ILT), constructed by ASTRON [van Haarlem et al. (2013)], during station-owners time. In this work we made use of data from the Effelsberg (DE601) LOFAR station funded by the Max-Planck-Gesellschaft; the Unterweilenbach (DE602) LOFAR station funded by the Max-Planck-Institut für Astrophysik, Garching; the Tautenburg (DE603) LOFAR station funded by the State of Thuringia, supported by the European Union (EFRE) and the Federal Ministry of Education and Research (BMBF) Verbundforschung project D-LOFAR I (grant 05A08ST1); the Potsdam (DE604) LOFAR station funded by the Leibniz-Institut für Astrophysik, Potsdam; the Jülich (DE605) LOFAR station supported by the BMBF Verbundforschung project D-LOFAR I (grant 05A08LJ1); and the Norderstedt (DE609) LOFAR station funded by the BMBF Verbundforschung project D-LOFAR II (grant 05A11LJ1). The observations of the German LOFAR stations were carried out in the stand-alone GLOW mode (German LOng-Wavelength array), which is technically operated and supported by the Max-Planck-Institut für Radioastronomie, the Forschungszentrum Jülich and Bielefeld University. We acknowledge support and operation of the GLOW network, computing and storage facilities by the FZ-Jülich, the MPIfR and Bielefeld University and financial support from BMBF D-LOFAR III (grant 05A14PBA) and D-LOFAR IV (grant 05A17PBA), and by the states of Nordrhein-Westfalia and Hamburg.

\bibliographystyle{aa} 
\bibliography{bibliography/ref}

\begin{thebibliography}{60}
\expandafter\ifx\csname natexlab\endcsname\relax\def\natexlab#1{#1}\fi

\bibitem[{Agazie {et~al.}(2023)Agazie, Anumarlapudi, Archibald, Arzoumanian,
  Baker, Bécsy, Blecha, Brazier, Brook, Burke-Spolaor, Burnette, Case,
  Charisi, Chatterjee, Chatziioannou, Cheeseboro, Chen, Cohen, Cordes, Cornish,
  Crawford, Cromartie, Crowter, Cutler, DeCesar, DeGan, Demorest, Deng, Dolch,
  Drachler, Ellis, Ferrara, Fiore, Fonseca, Freedman, Garver-Daniels, Gentile,
  Gersbach, Glaser, Good, Gültekin, Hazboun, Hourihane, Islo, Jennings,
  Johnson, Jones, Kaiser, Kaplan, Kelley, Kerr, Key, Klein, Laal, Lam, Lamb,
  Lazio, Lewandowska, Littenberg, Liu, Lommen, Lorimer, Luo, Lynch, Ma,
  Madison, Mattson, McEwen, McKee, McLaughlin, McMann, Meyers, Meyers,
  Mingarelli, Mitridate, Natarajan, Ng, Nice, Ocker, Olum, Pennucci, Perera,
  Petrov, Pol, Radovan, Ransom, Ray, Romano, Sardesai, Schmiedekamp,
  Schmiedekamp, Schmitz, Schult, Shapiro-Albert, Siemens, Simon, Siwek, Stairs,
  Stinebring, Stovall, Sun, Susobhanan, Swiggum, Taylor, Taylor, Turner, Unal,
  Vallisneri, van Haasteren, Vigeland, Wahl, Wang, Witt, Young, \&
  Collaboration}]{nanogravgwb}
Agazie, G., {et~al.}, Archibald, A.~M., Arzoumanian, Z. 2023, The Astrophysical
  Journal Letters, 951, L8

\bibitem[{{Alpar} {et~al.}(1986){Alpar}, {Nandkumar}, \& {Pines}}]{alparptn}
{Alpar}, M.~A., {Nandkumar}, R., \& {Pines}, D. 1986, \apj, 311, 197

\bibitem[{Antoniadis {et~al.}(2023{\natexlab{a}})Antoniadis, Arumugam,
  Arumugam, Babak, Bagchi, Bak~Nielsen, Bassa, Bathula, Berthereau, Bonetti,
  Bortolas, Brook, Burgay, Caballero, Chalumeau, Champion, Chanlaridis, Chen,
  Cognard, Dandapat, Deb, Desai, Desvignes, Dhanda-Batra, Dwivedi, Falxa,
  Ferdman, Franchini, Gair, Goncharov, Gopakumar, Graikou, Grießmeier,
  Guillemot, Guo, Gupta, Hisano, Hu, Iraci, Izquierdo-Villalba, Jang, Jawor,
  Janssen, Jessner, Joshi, Kareem, Karuppusamy, Keane, Keith, Kharbanda,
  Kikunaga, Kolhe, Kramer, Krishnakumar, Lackeos, Lee, Liu, Liu, Lyne, McKee,
  Maan, Main, Mickaliger, Niţu, Nobleson, Paladi, Parthasarathy, Perera,
  Perrodin, Petiteau, Porayko, Possenti, Prabu, Quelquejay~Leclere, Rana,
  Samajdar, Sanidas, Sesana, Shaifullah, Singha, Speri, Spiewak, Srivastava,
  Stappers, Surnis, Susarla, Susobhanan, Takahashi, Tarafdar, Theureau,
  Tiburzi, van~der Wateren, Vecchio, Venkatraman~Krishnan, Verbiest, Wang,
  Wang, \& Wu}]{eptagwb}
Antoniadis, J., {et~al.}, Arumugam, S., Babak, S. 2023{\natexlab{a}}, Astronomy
  \&; Astrophysics, 678, A50

\bibitem[{Antoniadis {et~al.}(2023{\natexlab{b}})Antoniadis, Babak,
  Bak~Nielsen, Bassa, Berthereau, Bonetti, Bortolas, Brook, Burgay, Caballero,
  Chalumeau, Champion, Cłianlaridis, Chen, Cognard, Desvignes, Falxa, Ferdman,
  Franchini, Gair, Goncharov, Graikou, Grießmeier, Guillemot, Guo, Hu, Iraci,
  Izquierdo-Villalba, Jang, Jawor, Janssen, Jessner, Karuppusamy, Keane, Keith,
  Kramer, Krishnakumar, Lackeos, Lee, Liu, Liu, Lyne, McKee, Main, Mickaliger,
  Niţu, Parthasarathy, Perera, Perrodin, Petiteau, Porayko, Possenti,
  Quelquejay~Leclere, Samajdar, Sanidas, Sesana, Shaifullah, Speri, Spiewak,
  Stappers, Susarla, Theureau, Tiburzi, van~der Wateren, Vecchio,
  Venkatraman~Krishnan, Verbiest, Wang, Wang, \& Wu}]{eptadr2}
Antoniadis, J., {et~al.}, Bak~Nielsen, A.-S., Bassa, C.~G. 2023{\natexlab{b}},
  Astronomy \&; Astrophysics, 678, A48

\bibitem[{{Arzoumanian et al}(2018)}]{11yrnano}
{Arzoumanian et al}, Z. 2018, \apjs, 235, 37

\bibitem[{{Backer} {et~al.}(1982){Backer}, {Kulkarni}, {Heiles}, {Davis}, \&
  {Goss}}]{mspcitation}
{Backer}, D.~C., {et~al.}, {Heiles}, C., {Davis}, M.~M. 1982, \nat, 300, 615

\bibitem[{{Biermann}(1951)}]{Biermann1951}
{Biermann}, L. 1951, \zap, 29, 274

\bibitem[{Bray {et~al.}(2015)Bray, Ekers, Roberts, Reynolds, James, Phillips,
  Protheroe, McFadden, \& Aartsen}]{lunaska}
Bray, J., {et~al.}, Roberts, P., Reynolds, J. 2015, Astroparticle Physics, 65,
  22

\bibitem[{Chalumeau {et~al.}(2021)Chalumeau, Babak, Petiteau, Chen, Samajdar,
  Caballero, Theureau, Guillemot, Desvignes, Parthasarathy, Liu, Shaifullah,
  Hu, van der Wateren, Antoniadis, Bak Nielsen, Bassa, Berthereau, Burgay,
  Champion, Cognard, Falxa, Ferdman, Freire, Gair, Graikou, Guo, Jang, Janssen,
  Karuppusamy, Keith, Kramer, Lee, Liu, Lyne, Main, McKee, Mickaliger, Perera,
  Perrodin, Porayko, Possenti, Sanidas, Sesana, Speri, Stappers, Tiburzi,
  Vecchio, Verbiest, Wang, Wang, \& Xu}]{chalumeau2021}
Chalumeau, A., {et~al.}, Petiteau, A., Chen, S. 2021, Monthly Notices of the
  Royal Astronomical Society, 509, 5538

\bibitem[{Counselman~III \& Rankin(1972)}]{counselman1972}
Counselman~III, C. \& Rankin, J. 1972, Astrophysical Journal, vol. 175, p. 843,
  175, 843

\bibitem[{{Donner} {et~al.}(2020{\natexlab{a}}){Donner}, {Verbiest, J. P. W.},
  {Tiburzi, C.}, {Osłowski, S.}, {Künsemöller, J.}, {Bak Nielsen, A.-S.},
  {Grießmeier, J.-M.}, {Serylak, M.}, {Kramer, M.}, {Anderson, J. M.},
  {Wucknitz, O.}, {Keane, E.}, {Kondratiev, V.}, {Sobey, C.}, {McKee, J. W.},
  {Bilous, A. V.}, {Breton, R. P.}, {Brüggen, M.}, {Ciardi, B.}, {Hoeft, M.},
  {van Leeuwen, J.}, \& {Vocks, C.}}]{donner2020}
{Donner}, {et~al.}, {Tiburzi, C.}, {Osłowski, S.} 2020{\natexlab{a}}, A\& A,
  644, A153

\bibitem[{{Donner} {et~al.}(2020{\natexlab{b}}){Donner}, {Verbiest}, {Tiburzi},
  {Os{\l}owski}, {K{\"u}nsem{\"o}ller}, {Bak Nielsen}, {Grie{\ss}meier},
  {Serylak}, {Kramer}, {Anderson}, {Wucknitz}, {Keane}, {Kondratiev}, {Sobey},
  {McKee}, {Bilous}, {Breton}, {Br{\"u}ggen}, {Ciardi}, {Hoeft}, {van Leeuwen},
  \& {Vocks}}]{Donner202036msp}
{Donner}, J.~Y., {et~al.}, {Tiburzi}, C., {Os{\l}owski}, S. 2020{\natexlab{b}},
  \aap, 644, A153

\bibitem[{Edwards {et~al.}(2006)Edwards, Hobbs, \& Manchester}]{Hobbsetal2006b}
Edwards, R.~T., Hobbs, G.~B., \& Manchester, R.~N. 2006, Monthly Notices of the
  Royal Astronomical Society, 372, 1549

\bibitem[{{Ellis} {et~al.}(2019){Ellis}, {Vallisneri}, {Taylor}, \&
  {Baker}}]{enterpriseellis}
{Ellis}, J.~A., {Vallisneri}, M., {Taylor}, S.~R., \& {Baker}, P.~T. 2019,
  {ENTERPRISE: Enhanced Numerical Toolbox Enabling a Robust PulsaR Inference
  SuitE}, Astrophysics Source Code Library, record ascl:1912.015

\bibitem[{{EPTA Collaboration} {et~al.}(2023){EPTA Collaboration}, {InPTA
  Collaboration}, {Antoniadis}, {Arumugam}, {Arumugam}, {Babak}, {Bagchi},
  {Nielsen}, {Bassa}, {Bathula}, {Berthereau}, {Bonetti}, {Bortolas}, {Brook},
  {Burgay}, {Caballero}, {Chalumeau}, {Champion}, {Chanlaridis}, {Chen},
  {Cognard}, {Dandapat}, {Deb}, {Desai}, {Desvignes}, {Dhanda-Batra},
  {Dwivedi}, {Falxa}, {Ferdman}, {Franchini}, {Gair}, {Goncharov}, {Gopakumar},
  {Graikou}, {Grie{\ss}meier}, {Guillemot}, {Guo}, {Gupta}, {Hisano}, {Hu},
  {Iraci}, {Izquierdo-Villalba}, {Jang}, {Jawor}, {Janssen}, {Jessner},
  {Joshi}, {Kareem}, {Karuppusamy}, {Keane}, {Keith}, {Kharbanda}, {Kikunaga},
  {Kolhe}, {Kramer}, {Krishnakumar}, {Lackeos}, {Lee}, {Liu}, {Liu}, {Lyne},
  {McKee}, {Maan}, {Main}, {Mickaliger}, {Ni{\c{t}}u}, {Nobleson}, {Paladi},
  {Parthasarathy}, {Perera}, {Perrodin}, {Petiteau}, {Porayko}, {Possenti},
  {Prabu}, {Leclere}, {Rana}, {Samajdar}, {Sanidas}, {Sesana}, {Shaifullah},
  {Singha}, {Speri}, {Spiewak}, {Srivastava}, {Stappers}, {Surnis}, {Susarla},
  {Susobhanan}, {Takahashi}, {Tarafdar}, {Theureau}, {Tiburzi}, {van der
  Wateren}, {Vecchio}, {Krishnan}, {Verbiest}, {Wang}, {Wang}, \&
  {Wu}}]{eptawm2}
{EPTA Collaboration}, {et~al.}, {Antoniadis}, J., {Arumugam}, P. 2023, \aap,
  678, A49

\bibitem[{Feldman {et~al.}(1998)Feldman, Sch{\"u}hle, Widing, \&
  Laming}]{Feldman1998}
Feldman, U., Sch{\"u}hle, U., Widing, K.~G., \& Laming, J.~M. 1998, The
  Astrophysical Journal, 505, 999

\bibitem[{Goncharov {et~al.}(2020)Goncharov, Reardon, Shannon, Zhu, Thrane,
  Bailes, Bhat, Dai, Hobbs, Kerr, Manchester, Osłowski, Parthasarathy,
  Russell, Spiewak, Thyagarajan, \& Wang}]{goncharov2021}
Goncharov, B., {et~al.}, Shannon, R.~M., Zhu, X.-J. 2020, Monthly Notices of
  the Royal Astronomical Society, 502, 478

\bibitem[{{Guhathakurta} \& {Fisher}(1998)}]{guhathakurta}
{Guhathakurta}, M. \& {Fisher}, R. 1998, \apjl, 499, L215

\bibitem[{Hazboun(2020)}]{la-forge}
Hazboun, J.~S. 2020, La Forge

\bibitem[{{Hazboun} {et~al.}(2020){Hazboun}, {Simon}, {Taylor}, {Lam},
  {Vigeland}, {Islo}, {Key}, {Arzoumanian}, {Baker}, {Brazier}, {Brook},
  {Burke-Spolaor}, {Chatterjee}, {Cordes}, {Cornish}, {Crawford}, {Crowter},
  {Cromartie}, {DeCesar}, {Demorest}, {Dolch}, {Ellis}, {Ferdman}, {Ferrara},
  {Fonseca}, {Garver-Daniels}, {Gentile}, {Good}, {Holgado}, {Huerta},
  {Jennings}, {Jones}, {Jones}, {Kaiser}, {Kaplan}, {Kelley}, {Lazio}, {Levin},
  {Lommen}, {Lorimer}, {Luo}, {Lynch}, {Madison}, {McLaughlin}, {McWilliams},
  {Mingarelli}, {Ng}, {Nice}, {Pennucci}, {Pol}, {Ransom}, {Ray}, {Siemens},
  {Spiewak}, {Stairs}, {Stinebring}, {Stovall}, {Swiggum}, {Turner},
  {Vallisneri}, {van Haasteren}, {Witt}, \& {Zhu}}]{hazboun2020}
{Hazboun}, J.~S., {et~al.}, {Taylor}, S.~R., {Lam}, M.~T. 2020, \apj, 890, 108

\bibitem[{{Hazboun et al}(2022)}]{Hazboun2022}
{Hazboun et al}, J. 2022, The Astrophysical Journal, 929, 39

\bibitem[{{Hellings} \& {Downs}(1983)}]{hd1983}
{Hellings}, R.~W. \& {Downs}, G.~S. 1983, \apjl, 265, L39

\bibitem[{{Hobbs} {et~al.}(2006){Hobbs}, {Lyne}, \& {Kramer}}]{Hobbs06ptn}
{Hobbs}, G., {Lyne}, A., \& {Kramer}, M. 2006, Chinese Journal of Astronomy and
  Astrophysics Supplement, 6, 169

\bibitem[{Hobbs {et~al.}(2006)Hobbs, Edwards, \& Manchester}]{Hobbsetal2006a}
Hobbs, G.~B., Edwards, R.~T., \& Manchester, R.~N. 2006, Monthly Notices of the
  Royal Astronomical Society, 369, 655

\bibitem[{{Hotan} {et~al.}(2004){Hotan}, {van Straten}, \&
  {Manchester}}]{hotanpsrchive}
{Hotan}, A.~W., {van Straten}, W., \& {Manchester}, R.~N. 2004, \pasa, 21, 302

\bibitem[{Issautier {et~al.}(2003)Issautier, Moncuquet, \& Hoang}]{ulysses}
Issautier, K., Moncuquet, M., \& Hoang, S. 2003, AIP Conference Proceedings,
  679, 59

\bibitem[{{Keith} {et~al.}(2013){Keith}, {Coles}, {Shannon}, {Hobbs},
  {Manchester}, {Bailes}, {Bhat}, {Burke-Spolaor}, {Champion}, {Chaudhary},
  {Hotan}, {Khoo}, {Kocz}, {Os{\l}owski}, {Ravi}, {Reynolds}, {Sarkissian},
  {van Straten}, \& {Yardley}}]{keith2013}
{Keith}, M.~J., {et~al.}, {Shannon}, R.~M., {Hobbs}, G.~B. 2013, \mnras, 429,
  2161

\bibitem[{{Kulkarni}(2020)}]{kulkarni2020}
{Kulkarni}, S.~R. 2020, arXiv e-prints, arXiv:2007.02886

\bibitem[{Kumar {et~al.}(2022)Kumar, White, Stovall, Dowell, \&
  Taylor}]{kumar2022}
Kumar, P., {et~al.}, Stovall, K., Dowell, J. 2022, Monthly Notices of the Royal
  Astronomical Society, 511, 3937

\bibitem[{{Lazarus} {et~al.}(2016){Lazarus}, {Karuppusamy}, {Graikou},
  {Caballero}, {Champion}, {Lee}, {Verbiest}, \& {Kramer}}]{lazaruscoast}
{Lazarus}, P., {et~al.}, {Graikou}, E., {Caballero}, R.~N. 2016, \mnras, 458,
  868

\bibitem[{{Liu} {et~al.}(2012){Liu}, {Keane}, {Lee}, {Kramer}, {Cordes}, \&
  {Purver}}]{liujitter}
{Liu}, K., {et~al.}, {Lee}, K.~J., {Kramer}, M. 2012, \mnras, 420, 361

\bibitem[{{Lorimer} \& {Kramer}(2004)}]{LorimerandKramer2005}
{Lorimer}, D.~R. \& {Kramer}, M. 2004, {Handbook of Pulsar Astronomy}, Vol.~4

\bibitem[{{{Madison} et~al.}(2019)}]{madison2019}
{{Madison} et~al.} 2019, \apj, 872, 150

\bibitem[{{Marsden} \& {Wenzel}(1991)}]{marsdenulysses}
{Marsden}, R.~G. \& {Wenzel}, K.~P. 1991, ESA Bulletin, 67, 78

\bibitem[{{McComas} {et~al.}(1998){McComas}, {Bame}, {Barker}, {Feldman},
  {Phillips}, {Riley}, \& {Griffee}}]{mccomas1998}
{McComas}, D.~J., {et~al.}, {Barker}, P., {Feldman}, W.~C. 1998, \ssr, 86, 563

\bibitem[{Melatos \& Link(2013)}]{superflu}
Melatos, A. \& Link, B. 2013, Monthly Notices of the Royal Astronomical
  Society, 437, 21

\bibitem[{{Muhleman} \& {Anderson}(1981)}]{muhlemann}
{Muhleman}, D.~O. \& {Anderson}, J.~D. 1981, \apj, 247, 1093

\bibitem[{Niţu {et~al.}(2024)Niţu, Keith, Tiburzi, Brüggen, Champion, Chen,
  Cognard, Desvignes, Dettmar, Grießmeier, Guillemot, Guo, Hoeft, Hu, Jang,
  Janssen, Jawor, Karuppusamy, Keane, Kramer, Künsemöller, Lackeos, Liu,
  Main, McKee, Porayko, Shaifullah, Theureau, \& Vocks}]{luliana2024}
Niţu, I.~C., {et~al.}, Tiburzi, C., Brüggen, M. 2024, Monthly Notices of the
  Royal Astronomical Society, 528, 3304

\bibitem[{{Papitashvili} {et~al.}(2014){Papitashvili}, {Bilitza}, \&
  {King}}]{omni}
{Papitashvili}, N., {Bilitza}, D., \& {King}, J. 2014, in 40th COSPAR
  Scientific Assembly, Vol.~40, C0.1--12--14

\bibitem[{Poletto(2013)}]{poletto2013}
Poletto, G. 2013, J Adv Res, 4, 215

\bibitem[{{Rasmussen} \& {Williams}(2006)}]{rasmussen}
{Rasmussen}, C.~E. \& {Williams}, C. K.~I. 2006, {Gaussian Processes for
  Machine Learning}

\bibitem[{Reardon {et~al.}(2023)Reardon, Zic, Shannon, Hobbs, Bailes, Marco,
  Kapur, Rogers, Thrane, Askew, Bhat, Cameron, Curyło, Coles, Dai, Goncharov,
  Kerr, Kulkarni, Levin, Lower, Manchester, Mandow, Miles, Nathan, Osłowski,
  Russell, Spiewak, Zhang, \& Zhu}]{pptagwb}
Reardon, D.~J., {et~al.}, Shannon, R.~M., Hobbs, G.~B. 2023, The Astrophysical
  Journal Letters, 951, L6

\bibitem[{{Shaifullah} {et~al.}(2023){Shaifullah}, {Magdalenic}, {Tiburzi},
  {Jebaraj}, {Samara}, \& {Zucca}}]{smt23}
{Shaifullah}, G.~M., {et~al.}, {Tiburzi}, C., {Jebaraj}, I. 2023, Advances in
  Space Research, 72, 5298

\bibitem[{{Swiggum} \& {NANOGrav Pfc}(2022)}]{nanograv15y}
{Swiggum}, J. \& {NANOGrav Pfc}. 2022, in American Astronomical Society Meeting
  Abstracts, Vol.~54, American Astronomical Society Meeting \#240, 348.08

\bibitem[{Tarafdar {et~al.}(2022)Tarafdar, Nobleson, Rana, Singha,
  Krishnakumar, Joshi, Paladi, Kolhe, Batra, Agarwal, Bathula, Dandapat, Desai,
  Dey, Hisano, Ingale, Kato, Kharbanda, Kikunaga, Marmat, Pandian, Prabu,
  Srivastava, Surnis, Susarla, Susobhanan, Takahashi, Arumugam, Bagchi, Banik,
  De, Girgaonkar, Gopakumar, Gupta, Maan, Manoharan, Naidu, \&
  Pathak}]{inptadr1}
Tarafdar, P., {et~al.}, Rana, P., Singha, J. 2022, Publications of the
  Astronomical Society of Australia, 39

\bibitem[{Taylor {et~al.}(2021)Taylor, Baker, Hazboun, Simon, \&
  Vigeland}]{enterprise}
Taylor, S.~R., {et~al.}, Hazboun, J.~S., Simon, J. 2021,
  enterprise\_extensions, v2.3.3

\bibitem[{Tiburzi {et~al.}(2015)Tiburzi, Hobbs, Kerr, Coles, Dai, Manchester,
  Possenti, Shannon, \& You}]{Tiburzi2015}
Tiburzi, C., {et~al.}, Kerr, M., Coles, W.~A. 2015, Monthly Notices of the
  Royal Astronomical Society, 455, 4339

\bibitem[{{Tiburzi} {et~al.}(2023){Tiburzi}, {Jackson}, {Cota}, {Shaifullah},
  {Fallows}, {Tokumaru}, \& {Zucca}}]{tjc23}
{Tiburzi}, C., {et~al.}, {Cota}, L., {Shaifullah}, G.~M. 2023, Advances in
  Space Research, 72, 5287

\bibitem[{{Tiburzi} {et~al.}(2021){Tiburzi}, {Shaifullah}, {Bassa}, {Zucca},
  {Verbiest}, {Porayko}, {van der Wateren}, {Fallows}, {Main}, {Janssen},
  {Anderson}, {Bak Nielsen}, {Donner}, {Keane}, {K{\"u}nsem{\"o}ller},
  {Os{\l}owski}, {Grie{\ss}meier}, {Serylak}, {Br{\"u}ggen}, {Ciardi},
  {Dettmar}, {Hoeft}, {Kramer}, {Mann}, \& {Vocks}}]{Tiburzi2021}
{Tiburzi}, C., {et~al.}, {Bassa}, C.~G., {Zucca}, P. 2021, \aap, 647, A84

\bibitem[{Tiburzi {et~al.}(2019)Tiburzi, Verbiest, Shaifullah, Janssen,
  Anderson, Horneffer, Künsemöller, Osłowski, Donner, Kramer, Kumari,
  Porayko, Zucca, Ciardi, Dettmar, Grießmeier, Hoeft, \&
  Serylak}]{tiburzi2019}
Tiburzi, C., {et~al.}, Shaifullah, G.~M., Janssen, G.~H. 2019, Monthly Notices
  of the Royal Astronomical Society, 487, 394

\bibitem[{{Tokumaru} {et~al.}(2020){Tokumaru}, {Tawara}, {Takefuji}, {Sekido},
  \& {Terasawa}}]{Tokumaru2020}
{Tokumaru}, M., {et~al.}, {Takefuji}, K., {Sekido}, M. 2020, \solphys, 295, 80

\bibitem[{{van Haarlem} {et~al.}(2013){van Haarlem}, {Wise}, {Gunst}, {Heald},
  {McKean}, {Hessels}, {de Bruyn}, {Nijboer}, {Swinbank}, {Fallows},
  {Brentjens}, {Nelles}, {Beck}, {Falcke}, {Fender}, {H{\"o}randel},
  {Koopmans}, {Mann}, {Miley}, {R{\"o}ttgering}, {Stappers}, {Wijers},
  {Zaroubi}, {van den Akker}, {Alexov}, {Anderson}, {Anderson}, {van Ardenne},
  {Arts}, {Asgekar}, {Avruch}, {Batejat}, {B{\"a}hren}, {Bell}, {Bell}, {van
  Bemmel}, {Bennema}, {Bentum}, {Bernardi}, {Best}, {B{\^\i}rzan}, {Bonafede},
  {Boonstra}, {Braun}, {Bregman}, {Breitling}, {van de Brink}, {Broderick},
  {Broekema}, {Brouw}, {Br{\"u}ggen}, {Butcher}, {van Cappellen}, {Ciardi},
  {Coenen}, {Conway}, {Coolen}, {Corstanje}, {Damstra}, {Davies}, {Deller},
  {Dettmar}, {van Diepen}, {Dijkstra}, {Donker}, {Doorduin}, {Dromer}, {Drost},
  {van Duin}, {Eisl{\"o}ffel}, {van Enst}, {Ferrari}, {Frieswijk}, {Gankema},
  {Garrett}, {de Gasperin}, {Gerbers}, {de Geus}, {Grie{\ss}meier}, {Grit},
  {Gruppen}, {Hamaker}, {Hassall}, {Hoeft}, {Holties}, {Horneffer}, {van der
  Horst}, {van Houwelingen}, {Huijgen}, {Iacobelli}, {Intema}, {Jackson},
  {Jelic}, {de Jong}, {Juette}, {Kant}, {Karastergiou}, {Koers}, {Kollen},
  {Kondratiev}, {Kooistra}, {Koopman}, {Koster}, {Kuniyoshi}, {Kramer},
  {Kuper}, {Lambropoulos}, {Law}, {van Leeuwen}, {Lemaitre}, {Loose}, {Maat},
  {Macario}, {Markoff}, {Masters}, {McFadden}, {McKay-Bukowski}, {Meijering},
  {Meulman}, {Mevius}, {Middelberg}, {Millenaar}, {Miller-Jones}, {Mohan},
  {Mol}, {Morawietz}, {Morganti}, {Mulcahy}, {Mulder}, {Munk}, {Nieuwenhuis},
  {van Nieuwpoort}, {Noordam}, {Norden}, {Noutsos}, {Offringa}, {Olofsson},
  {Omar}, {Orr{\'u}}, {Overeem}, {Paas}, {Pandey-Pommier}, {Pandey}, {Pizzo},
  {Polatidis}, {Rafferty}, {Rawlings}, {Reich}, {de Reijer}, {Reitsma},
  {Renting}, {Riemers}, {Rol}, {Romein}, {Roosjen}, {Ruiter}, {Scaife}, {van
  der Schaaf}, {Scheers}, {Schellart}, {Schoenmakers}, {Schoonderbeek},
  {Serylak}, {Shulevski}, {Sluman}, {Smirnov}, {Sobey}, {Spreeuw}, {Steinmetz},
  {Sterks}, {Stiepel}, {Stuurwold}, {Tagger}, {Tang}, {Tasse}, {Thomas},
  {Thoudam}, {Toribio}, {van der Tol}, {Usov}, {van Veelen}, {van der Veen},
  {ter Veen}, {Verbiest}, {Vermeulen}, {Vermaas}, {Vocks}, {Vogt}, {de Vos},
  {van der Wal}, {van Weeren}, {Weggemans}, {Weltevrede}, {White}, {Wijnholds},
  {Wilhelmsson}, {Wucknitz}, {Yatawatta}, {Zarka}, {Zensus}, \& {van
  Zwieten}}]{vanhaarlem2013}
{van Haarlem}, M.~P., {et~al.}, {Gunst}, A.~W., {Heald}, G. 2013, \aap, 556, A2

\bibitem[{{van Haasteren} {et~al.}(2009){van Haasteren}, {Levin}, {McDonald},
  \& {Lu}}]{vanhaasteren2009}
{van Haasteren}, R., {Levin}, Y., {McDonald}, P., \& {Lu}, T. 2009, \mnras,
  395, 1005

\bibitem[{Van~Haasteren \& Vallisneri(2014)}]{van_Haasteren_2014}
Van~Haasteren, R. \& Vallisneri, M. 2014, Physical Review D, 90

\bibitem[{{van Straten} \& {Bailes}(2011)}]{vanstraaten2011}
{van Straten}, W. \& {Bailes}, M. 2011, \pasa, 28, 1

\bibitem[{{van Straten} {et~al.}(2011){van Straten}, {Demorest}, {Khoo},
  {Keith}, {Hotan}, \& {et al.}}]{straatenpsrchive}
{van Straten}, W., {et~al.}, {Khoo}, J., {Keith}, M. 2011, {PSRCHIVE:
  Development Library for the Analysis of Pulsar Astronomical Data},
  Astrophysics Source Code Library, record ascl:1105.014

\bibitem[{Wang(2015)}]{wang2015}
Wang, Y. 2015, Journal of Physics: Conference Series, 610, 012019

\bibitem[{You {et~al.}(2007{\natexlab{a}})You, Hobbs, Coles, Manchester,
  Edwards, Bailes, Sarkissian, Verbiest, Van~Straten, Hotan, Ord, Jenet, Bhat,
  \& Teoh}]{you2007dm}
You, X.~P., {et~al.}, Coles, W.~A., Manchester, R.~N. 2007{\natexlab{a}},
  Monthly Notices of the Royal Astronomical Society, 378, 493

\bibitem[{You {et~al.}(2007{\natexlab{b}})You, Hobbs, Coles, Manchester, \&
  Han}]{You2007}
You, X.~P., {et~al.}, Coles, W.~A., Manchester, R.~N. 2007{\natexlab{b}}, ApJ,
  671, 907

\bibitem[{Zic {et~al.}(2023)Zic, Reardon, Kapur, Hobbs, Mandow, Curyło,
  Shannon, Askew, Bailes, Bhat, Cameron, Chen, Dai, Marco, Feng, Kerr,
  Kulkarni, Lower, Luo, Manchester, Miles, Nathan, Osłowski, Rogers, Russell,
  Sarkissian, Shamohammadi, Spiewak, Thyagarajan, Toomey, Wang, Zhang, Zhang,
  \& Zhu}]{pptadr3}
Zic, A., {et~al.}, Kapur, A., Hobbs, G. 2023, The Parkes Pulsar Timing Array
  Third Data Release

\end{thebibliography}

\end{document}